\documentclass[a4paper,fleqn,usenatbib]{mnras}

\usepackage[T1]{fontenc}
\usepackage{ae,aecompl}

\usepackage{graphicx}	
\usepackage{amsmath}	
\usepackage{amssymb}	
\usepackage{ulem}       
\usepackage{cancel}     
\usepackage{float}
\usepackage{hyperref}
\usepackage{xcolor}
\usepackage{afterpage}
\usepackage{subcaption}
\usepackage{caption}

\def\orcid#1{\kern .08em\href{https://orcid.org/#1}{\includegraphics[keepaspectratio,width=0.7em]{ORCIDiD_icon16x16.pdf}}}

\title[Stream-Disk Interaction in MADs]{Jet Tilt Instability from Stream-Disk Interactions in MAD Disks}
\author[Curd et al.]{
Brandon Curd,$^{1,2,3}$\thanks{E-mail: brandon.curd@utsa.edu} Richard Anantua$^{1,2,3
}$, Hayley West$^{1}$, and Joaquin Duran$^{1}$
\\ $^{1}$ Department of Physics $\&$ Astronomy, The University of Texas at San Antonio, One UTSA Circle, San Antonio, TX 78249, USA
\\ $^{2}$ Black Hole Initiative at Harvard University, 20 Garden Street, Cambridge, MA 02138, USA
\\ $^{3}$ Center for Astrophysics $\vert$ Harvard \& Smithsonian, 60 Garden Street, Cambridge, MA 02138, USA
}

\date{Accepted XXX. Received YYY; in original form ZZZ}

\pubyear{2024}

\begin{document}
\label{firstpage}
\pagerange{\pageref{firstpage}--\pageref{lastpage}}
\maketitle

\begin{abstract}
Magnetically arrested accretion disks (MADs) around a rapidly rotating black hole (BH) have been proposed as a model for jetted tidal disruption events (TDEs). However, the stream and disk interact strongly at times, and this will lead to different dynamics than expected in the standard MAD model. Here we employ global GRMHD simulations of a MAD disk interacting with an injected stream with a penetrating pericenter $R_p\sim 10 r_g$ and a range of density contrasts $f_\rho\equiv \rho_d/\rho_s$, or how dense the disk is relative to the stream. We demonstrate for the first time that a MAD or semi-MAD state can be sustained and jets powered by the BH spin can be produced even when the stream is much denser than the disk, i.e. in the first month(s) of a jetted TDE. We also demonstrate that the strength of the self-intersection shock decreases as $f_\rho$, and time, increases. The jet or funnel can become significantly tilted (by $10-30^\circ$) due to the self-intersection outflow when $f_\rho \leq 0.1$. In models with a powerful jet and $f_\rho\leq 0.01$,  the tilted jet interacts with and ultimately tilts the disk by as much as 23 degrees from the incoming stream and this tilted state is stable for the duration of the simulation. As $f_\rho$ increases, the tilt of the jet and disk is expected to realign with the BH spin once $f_\rho \geq 0.1$. The jet tilt could rapidly realign due to outer disk collapse or the self-intersection radius increasing. Our results provide an alternative explanation for the observed X-ray jet shut-off in days-weeks in jetted TDEs.
\end{abstract}




\section{Introduction}
When a star wanders too close to its central black hole (BH), the tidal forces from the BH exceed the self gravity of the star and the star is subsequently disrupted into a stream of stellar material \citep{Hills1975,Rees1988,Phinney1989,Evans1989}. The bound portion of the stream ultimately returns to the BH, delivering mass to the periceneter radius at the fall back rate ($\dot{M}_{\rm{fb}}$) which falls off as $(t/t_{\rm{fb}})^{-5/3}$, where $t_{\rm{fb}}$ is the orbital period of the most bound portion of the stream (or the fall back time). This leads to emission which also drops off as $(t/t_{\rm{fb}})^{-5/3}$ since the energy available for dissipation is provided by the kinetic energy of the stream. The transient, which is known as a tidal disruption event (TDE), is typically detectable for months-years. 

The dynamics governing the properties of the stream and subsequent emission depend on the stellar mass, eccentricity, pericenter radius, and compressiblity of the star. The tidal radius of the star is given by,
\begin{equation} \label{eq:eq1}
  R_t/r_g = 47 m_6^{-2/3} m_*^{-1/3}r_*,
\end{equation}
\citep{Rees1988} where $m_6=M_{\rm{BH}}/10^6\,M_\odot$ is the mass of the SMBH, $m_*=M_{\rm{*}}/M_\odot$ is the mass of the disrupted star, and $r_*=R_{\rm{*}}/R_\odot$ is its radius. For the typical TDE, the orbit is parabolic ($e=1$). For zero age main sequence stars the radius for complete disruption depends on the compressibility and occurs at $\sim0.9 R_t$ for $\gamma=5/3$ and at $\gtrsim 2 R_t$ for $\gamma=4/3$ \citep{Guillochon2013,Mainetti2017}, though it is larger for evolved stars \citep{Golightly2019}. Several works have addressed the initial disruption of the star and evolution of the stream over a broad parameter space \citep{Carter1982,Evans1989,Kochanek1994,Lodato2009,Brassart2010,Stone2013,Coughlin2015,Coughlin2016,Steinberg2019,Ryu2020}. 

TDEs have been discovered in the X-ray, optical/UV, and radio (see \citealt{Komossa2015,Alexander2020,Gezari2021} for a review). While disk formation is expected in TDEs, what powers the emission is still unclear, for instance either turbulent accretion or shocks could explain the emission at different stages in the evolution. The presence of outflows, possibly launched by an accretion disk \citep{Strubbe2009,Coughlin2014,Metzger2016}, has been inferred in many cases due to radio emission \citep{Alexander2016,Alexander2017} and TDEs have also been observed to launch relativistic jets \citep{Bloom2011,Burrows2011,Zauderer2011,Cenko2012,Brown2015,Pasham2015,Eftekhari2024}. More recently, a handful of TDEs have been observed during the rise to peak \citep{Holoien2019,Holoien2020,Hinkle2021,Hammerstein2023}. This bounty of observations is expected to grow significantly once the Large Synoptic Survey Telescope (LSST, \citealt{Ivezic2019, Bricman2020}) comes online, but theory has yet to fully describe the range of observational properties exhibited by TDEs.

Jetted TDEs have observational properties that present a particularly complicated puzzle. For instance, \textit{Swift} J1644+57 showed rapid viability following the turn on with quasi-periodic oscillations (QPOs) at $\sim200$ s \citep{Reis2012}, long period viability at $\sim10^6$ s with the period increasing over the course of the transient \citep{Saxton2012}, and a rapid drop in the X-ray flux at the $\sim500$ days after the initial trigger \citep{Zauderer2013}. A similar drop off in the X-ray flux was seen in \textit{Swift} J2058+05 and AT2022cmc after several months \citep{Pasham2015,Eftekhari2024}. Since the peak accretion rate in TDEs around BHs of mass $M_{\rm{BH}}\lesssim 2.6\times10^7M_\odot$ may be super-Eddington (\autoref{eq:mpeak}), it is generally thought that the shut-off in the X-ray flux is associated with the disk transitioning from super- to sub-Eddington.

Magnetically arrested accretion disks (MADs, \citealt{Narayan2003}) are thought to provide a physical explanation for both the presence of a relativistic jets and variability in jetted TDEs. The large magnetic flux required for a MAD is thought to be sourced by either poloidal field lines in a fossil disk \citep{Kelley2014,Tchekhovskoy2014,Teboul2023} or conversion of toroidal field lines to poloidal through a dynamo effect \citep{Liska2020}. In the model presented by \cite{Tchekhovskoy2014}, the X-ray shut-off in \textit{Swift} J1644+57 is triggered because the disk becomes geometrically thin as it enters sub-Eddington accretion rates and subsequently loses its magnetic flux, which is necessary to power the jet. However, general relativistic radiation magnetohydrodynamics (GRRMHD) simulations of thin MADs have not shown complete jet turn off, potentially due to magnetic pressure support of the disk at low accretion rates \citep{Avara2016,Curd2023,Liska2022}. Thus, the rapid shut-off in X-ray flux is difficult to explain in a MAD state unless simulations are unable to capture magnetic diffusion due to their relatively short duration (typically several days).   

Disk formation in TDEs may result in a different disk structure than the standard advection dominated accretion disk (ADAF, \citealt{Abramowicz1988,Narayan1995}), which has been assumed in some studies \citep{Dai2018,Curd2019}. Several numerical studies of disk formation have demonstrated the presence of shocks and outflows as well as long lasting asymmetric structure \citep{Ramirez-Ruiz2009,Guillochon2013,Shiokawa2015,Bonnerot2016, Sadowski2016, Hayasaki2016, Bonnerot2020,Bonnerot2021, Curd2021,Andalman2022,Ryu2023,Steinberg2024}. Furthermore, the eccentricity of material sourced from the stream is difficult to dissipate which, in the majority of studies, leads to an eccentric accretion disk and spiral shocks. For instance, the most realistic smooth particle hydrodynamics simulations to date found imperfect circularization as the final disk remains mildly eccentric with $e\approx 0.3$ \citep{Bonnerot2020,Bonnerot2021}. A long duration simulation ($2 t_{\rm fb}$) by \citet{Ryu2023} demonstrated that shocks may dominate the energy budget of the TDE and the debris cloud may remain highly eccentric with $e\sim0.5-0.6$. However, recent RHD simulations with adaptive mesh refinement find that the inner debris was able to reach $e<0.2$ after more than 30 days \citep{Steinberg2024}, which is substantially longer than disk formation simulations with similar parameters. A notable development in TDE theory from \cite{Steinberg2024} is the nomenclature regarding the circularized debris. The circularized material has typically been referred to as a `disk', but their work demonstrates that the material is more appropriately referred to as a debris cloud. It is worth noting that GRMHD and GRRMHD simulations were unable to reach the magnetic flux that is required for the MAD state due to the weak magnetic flux provided by the stream as well as the chaotic disk formation \citep{Sadowski2016, Curd2021}. As there are no current simulations of eccentric MADs nor TDE disk formation simulations which result in a MAD, it is unclear how MADs in TDEs may differ from the standard thick accretion disk.

The primary questions we address in this work are (i) whether or not jetted TDEs can maintain the magnetic flux required for the MAD state when the stream-disk interaction is strongest (i.e. near the peak flux), and (ii) how the stream-disk interaction in jetted TDEs effects the dynamics compared to the standard MAD. Of course, these questions require the assumption that jetted TDEs are in fact MAD, a problem which we do not address in this work. Although \citet{Kelley2014} demonstrated that the stream can trap enough magnetic flux for the system to become MAD \textit{if it reaches the BH horizon}, how much magnetic field threaded the BH can not be seen in their local simulations. Global simulations of the full TDE disk formation process with magnetic fields included are needed in order to conclusively determine if TDEs can become MAD and under what conditions. It is important to note that the self intersection outflow is quasi spherical; therefore, the force that it applies to the inner disk and jet is not symmetrical (e.g. \citealt{Jiang2016}). This suggests that the jet, during strong self intersection, will experience an asymmetric lateral force about the jet axis. One might expect strong perturbation of the jet, and potentially the disk due its interaction with the jet. This could lead to bent jets, as has been observed in various radio sources \citep[e.g.][]{Ryle1968,Owen1976}. Our simulations provide insight into the dynamics of jetted TDEs since we capture both the BH horizon and self-intersection from the stream.

In this work, we investigate MAD or strongly magnetized disks interacting with streams of gas injected on penetrating orbits in GRMHD using a novel approach to overcome the computational difficulties in simulating the large and small scale structures, as well as long time scales, required to follow the full disruption in a global simulation. We also study the effects of spin and use $a_*=0$ and $a_*=0.9$. Prior to stream injection, we initialize each simulation with a circularized, small scale MAD disk which has been evolved until accretion sets in (15,000 $GM/c^2$). We then inject a magnetized stream with a constant mass flux, giving an $f_\rho$ which is related to the time evolution in a TDE (see Section \ref{sec:densitycontrast}). We set the pericenter radius of the stream such that the self intersection radius is on the order of $50 r_g$, where $r_g$ is the gravitational radius (defined in Section \ref{sec:definitions}). Since GRMHD simulations are scale free, the most important parameter in our simulations is the ratio between the density of the pre-existing disk and injected stream (or the density contrast, which we define in Section \ref{sec:densitycontrast}). We study the disk and jet properties during the interaction between the debris cloud/disk and stream.

This paper is organized as follows. In Section \ref{sec:densitycontrast}, we discuss how the density contrast evolves in a simplified model of the TDE stream and accretion disk and illustrate potential consequences on the dynamics. In Section \ref{sec:nummethods}, we describe the numerical methods used to perform the GRMHD simulations. In Section \ref{sec:definitions}, we define calculations used to analyze the simulations. In Section \ref{sec:results}, we discuss core results and provide visualizations of each simulation. We discuss how our results relate to jetted TDEs in Section \ref{sec:discussion} and conclude in Section \ref{sec:conclusions}.

\section{Density Contrast in TDEs} \label{sec:densitycontrast}

Following \citet{Stone2013}, we define the fallback time as
\begin{equation} \label{eq:fallbacktime}
    t_{\rm fb} = 3.5 \times 10^6 {\rm sec} \ m_6^{1/2}m_*^{-1}r_*^{3/2}.
\end{equation}
Following the rise to peak, the mass fallback rate follows a power law 
\begin{equation} \label{eq:massfallback}
    \dot{M}_{\rm fb} \sim \dot{M}_{\rm peak} \left( \frac{t}{t_{\rm fb}}\right)^{-5/3},
\end{equation}
where
\begin{equation} \label{eq:mpeak}
    \frac{\dot{M}_{\rm peak}}{\dot{M}_{\rm Edd}} \sim 133 m_6^{-3/2} m_*^2 r_*^{-3/2}
\end{equation}
is the peak mass fallback rate in units of the Eddington mass accretion rate (defined later in \autoref{eq:mdotEdd}). Note we set $\eta = 0.1$, $k = 1$, and $n = 0$ in each of the expressions for simplicity such that there is no dependence on $\beta$. 

The simulations presented in this work demonstrate that the density contrast 
\begin{equation}
  f_\rho(t,r) = \dfrac{\rho_d(t,r)}{\rho_s(t,r)} , 
\end{equation}leads to different dynamics in a TDE, where $\rho_d$ is the mass density of the pre-existing debris cloud/disk and  $\rho_s$  that of the injected stream. Namely, the self-intersection outflow can be diminished if the stream's orbit is changed during its interaction with the debris cloud/disk. At the start of the TDE evolution, $f_\rho<1$. Even in the simulation presented by \citet{Steinberg2024}, the circularized debris cloud clearly remains less dense than the stream by roughly an order of magnitude. Depending on how the debris cloud/disk mass, scale, and geometry evolves, the quantity $f_\rho$ could conceivably exceed unity at late times. Here we discuss how evolution of $f_\rho$ could be relevant in TDEs.

To describe the stream, we assume that its density is related to the fallback rate by the expression
\begin{equation}
    \rho_s(t,r) = \dfrac{\dot{M}_{\rm{fb}}(t)}{\pi H_s(r)^2 v_s(r)},
\end{equation}
where $H_s$ is the stream height and $v_s\approx \sqrt{2GM_{\rm{BH}}/r}$ is the free-fall velocity, which is roughly the speed of the incoming stream. For simplicity, we assume the stream height takes the form $H_s(r)=(r/R_t)R_*$.

To approximate the evolution of the debris cloud/disk, we assume that $t\geq t_{\rm{fb}}$ such that the initial debris cloud/disk mass is $M_d(t=t_{\rm{fb}})=0.1M_*$. We then approximate the debris cloud/disk mass by accounting for mass accreted by the BH over time
\begin{equation}
    \dot{M}_d(t) = \dot{M}_{\rm{fb}}(t) - \dot{M}_{\rm{BH}}(t).
\end{equation}
Here we assume $\dot{M}_{\rm{BH}}=f_{\rm{acc}}\dot{M}_{\rm{fb}}$, and use a fiducial value of $f_{\rm{acc}}=0.1$. This assumption is motivated by \citet{Curd2021}, which found a mass accretion rate of $\sim10\%$ of the fallback rate. This assumption may not hold for long term evolution as the debris cloud/disk mass builds up (e.g. \citealt{Metzger2022}). The debris cloud/disk mass then evolves as
\begin{equation}
    M_d(t) = M_{d,0} + (1-f_{\rm{acc}})\int_{t_{\rm{fb}}}^t \dot{M}_{\rm{fb}}(t)dt
\end{equation}

We assume that the gas density follows a power-law with radius of $\rho_d(r,t)=\rho_{d,0}(t)(r/r_H)^{-1}$, where $r_H$ is the horizon radius and $\rho_{d,0}(t)$ is the maximum density of the debris cloud/disk at time $t$. This profile is appropriate for a jetted TDE assuming the MAD disk has properties similar to generic MADs \citep{Chaterjee2022}, but is also similar to that of the TDE debris cloud/disk in \citet{Andalman2022}. The density for a debris cloud/disk of outer radius $R_d$ is obtained by
\begin{equation} \label{eq:rhod}
    \rho_{d}(t,r) = \dfrac{M_d(t)}{2\pi r(R_d^2 - r_H^2)}
\end{equation}
Here we assume a spherical distribution at all mass accretion rates. At low accretion rates, the debris cloud may collapse into a disk geometry with scale-height $h_d$ which may have radial dependence. We set $\rho_d(r)=0$ for $r<r_H$ and $r>R_d$.

\begin{figure}
    \centering{}
	\includegraphics[width=\columnwidth]{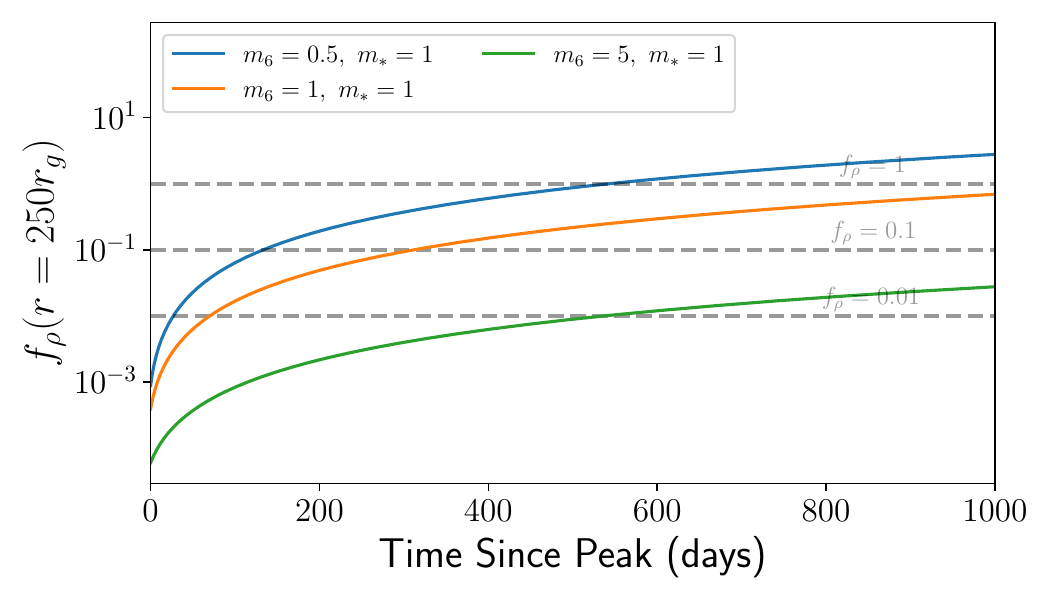}
    \caption{Here we illustrate how $f_\rho$ evolves in our simple TDE disk model with a BH mass $m_6=M_{\rm BH}/10^6 M_\odot$ and stellar mass $m_*=M_*/M_\odot$. Note that we set $R_d=500r_g$, $f_{\rm{acc}}=0.1$, and $\beta=1$ for simplicity. We show the initial $f_\rho$ (also measured at $r=250r_g$) for each $a_*=0.9$ simulation in Table \ref{tab:table1} based on $\dot{M}_{\rm{inj}}$ (horizontal dashed lines). As $f_\rho$ increases, the stream will dissipate more of its orbital energy in its interaction with the disk. As we describe in \autoref{sec:results}, the self intersection shock weakens as a result.}
    \label{fig:frhoex}
\end{figure}

Although we have performed simulations in which the accretion disk is geometrically thick, in part because we cannot sufficiently resolve small scale-height disks, our simulations demonstrate the impact that the density contrast has on the stream dynamics. The acceleration that the stream experiences as it interacts with the disk depends only on the density of the disk and the stream and is agnostic to the geometry of the disk. As such, we expect that the stream dynamics illustrated in our simulations should be similar in a geometrically thin system. Furthermore, the incoming stream is expected to be aligned with the disk since the debris tends to remain roughly aligned with the initial angular momentum of the star and does not precess \citep{Andalman2022}, so the stream is expected to strongly interact with the disk even if it becomes thin. However, it should be noted that the interaction between the self-intersection outflow and the disk material above/below the orbital plane may lead to different jet/outflow properties in thick versus thin disks.

We show examples of $f_\rho$ measured at $r=250r_g$ over time using our assumed debris cloud/disk and stream evolution in \autoref{fig:frhoex}. In a scenario where a circularized  debris cloud/accretion disk forms, there is not a cleared path for the stream to flow along towards pericenter. Instead, the circularized debris cloud/disk will exert ram pressure on the stream with an acceleration $a_{\rm ram}\propto f_\rho$, effectively braking it. At low $f_\rho$, the stream will be effectively unperturbed. However, as $f_\rho$ approaches unity, the ram pressure may completely prevent the stream from reaching pericenter. Instead, the stream may mix with the disk as it rapidly dissipates orbital energy similar to \cite{Steinberg2024}. Holding the other disruption parameters constant and assuming the same disk size, increasing (decreasing) the BH mass shifts the $f_\rho$ curve down (up), which changes where the transition from strong to weak self-intersection may occur. As we show in this work, the self intersection becomes weaker as $f_\rho$ increases, which leads to dynamic changes in the disk and jet/corona. Such evolution could be responsible for state transitions and delayed outflows, which have occurred in several TDEs. Here we have ignored the possibility of disk collapse, but we discuss how this may change TDE evolution in the context of $f_\rho$ in \autoref{sec:discussion}.

We note that the evolution, density profile, and size of the debris cloud/disk is a vital component of our asserted scenario. We have assumed a MAD profile, which peaks near the BH horizon, for consistency with our simulations and the \cite{Tchekhovskoy2014} model for jetted TDEs, but \cite{Steinberg2024} find that most of the mass is distributed at large radii. We have neglected any evolution in the size of the debris cloud/disk over time as we assumed $R_d$ is constant for simplicity, but radial contraction of the envelope will occur as the fallback rate declines \citep{Metzger2022}. In addition, we assume that $\dot{M}_{\rm{BH}}$ is proportional to $\dot{M}_{\rm{fb}}$ at all times. While this is based on simulation results, global simulations have yet to cover the full range of TDE evolution. In models such as \citet{Metzger2022}, bound material within the debris cloud/disk will also drain into the BH after an accretion time. See \citet{Metzger2022} for a description.

\section{Numerical Methods} \label{sec:nummethods}

We present a suite of 3D numerical simulations of MAD disks interacting with an injected stream carried out with the GRRMHD code, {\sc koral} \citep{Sadowski+2013,Sadowski+2014,Sadowski+2015a,Sadowski+2015b}. Using a mesh-based, finite-difference method in a stationary Kerr space-time, {\sc koral} solves the conservation equations of GRMHD: 
\begin{align}
  (\rho u^\mu)_{;\mu} &= 0, \label{eq:consrho} \\
  (T^\mu_{\ \nu})_{;\mu} &= 0, \label{eq:consT}
\end{align}
where $\rho$ is the gas density in the comoving fluid frame, $u^\mu$ are the components of the gas four-velocity as measured in the ``lab frame'', $T^\mu_{\ \nu}$ is the MHD stress-energy tensor in the ``lab frame'':
\begin{equation} \label{eq:Tmunu}
  T^\mu_{\ \nu} = (\rho + u_g+ p_g + b^2)u^\mu u_\nu + (p_g + \dfrac{1}{2}b^2)\delta^\mu_{\ \nu} - b^\mu b_\nu.
\end{equation}
Here $u_g$ and $p_g=(\gamma_g - 1)u_g$ are the internal energy and pressure of the gas in the comoving frame, and $b^\mu$ is the magnetic field four-vector which is evolved following the ideal MHD induction equation \citep{Gammie+2003}. We adopt $\gamma=5/3$ in this work. The code can handle radiation as well, but we choose to study pure GRMHD in this work to lower computational costs.

We evolve the fluid in modified Kerr-Schild coordinates with the inner radius of the simulation domain inside of the BH horizon. The radial grid cells are spaced logarithmically, and we choose inner and outer radial bounds $R_{\rm{min}}<r_H$ (with 4 cells within the horizon) and $R_{\rm{max}}=5\times 10^4\,r_g$. We also use a full $2\pi$ in azimuth and set $\varphi_{\rm{min}}=-\pi$ and $\varphi_{\rm{max}}=\pi$. We choose outflow boundary conditions at both the inner and outer radial bounds, reflective boundary conditions at the top and bottom polar boundaries, and periodic boundary conditions in $\varphi$. In each simulation, we employ a resolution $N_r\times N_\vartheta\times N_\varphi=256\times144\times144$. Specifics of the grid are given in Appendix \ref{sec:appB}.

\begin{figure}
    \centering{}
	\includegraphics[width=\columnwidth]{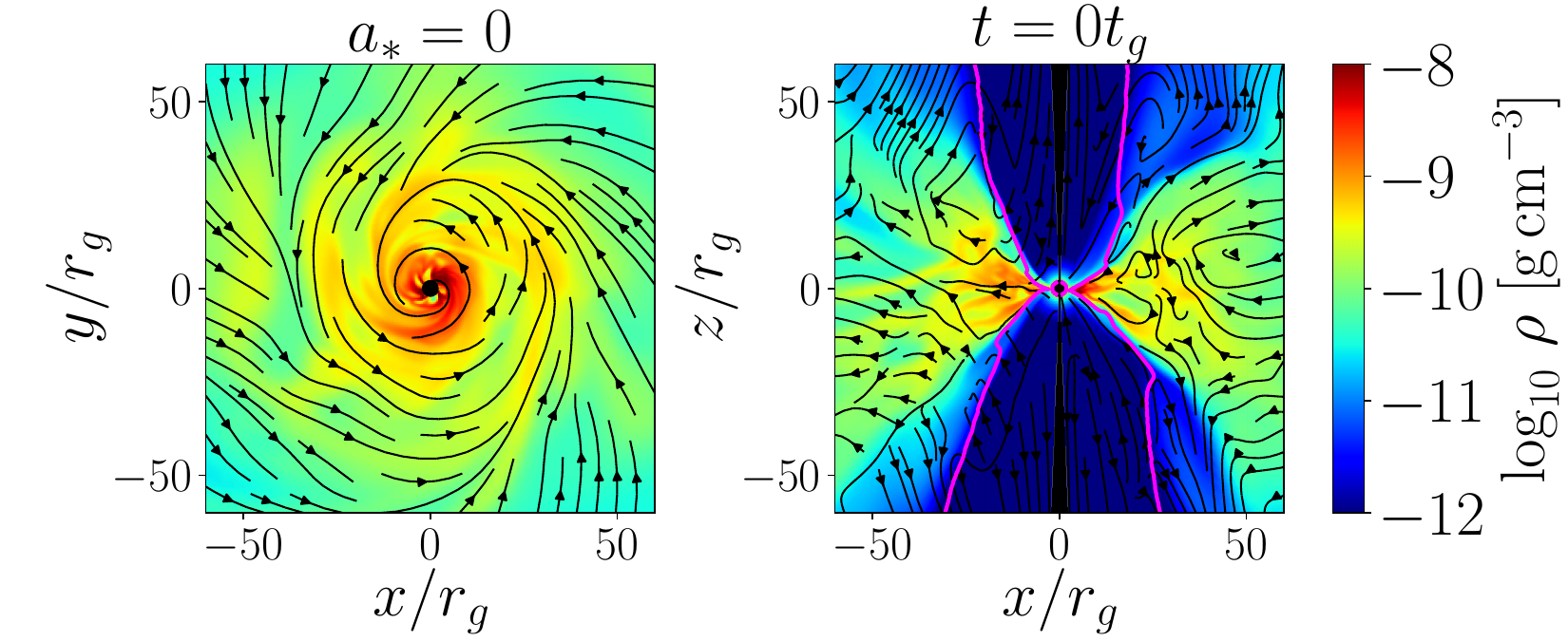}\\
	\includegraphics[width=\columnwidth]{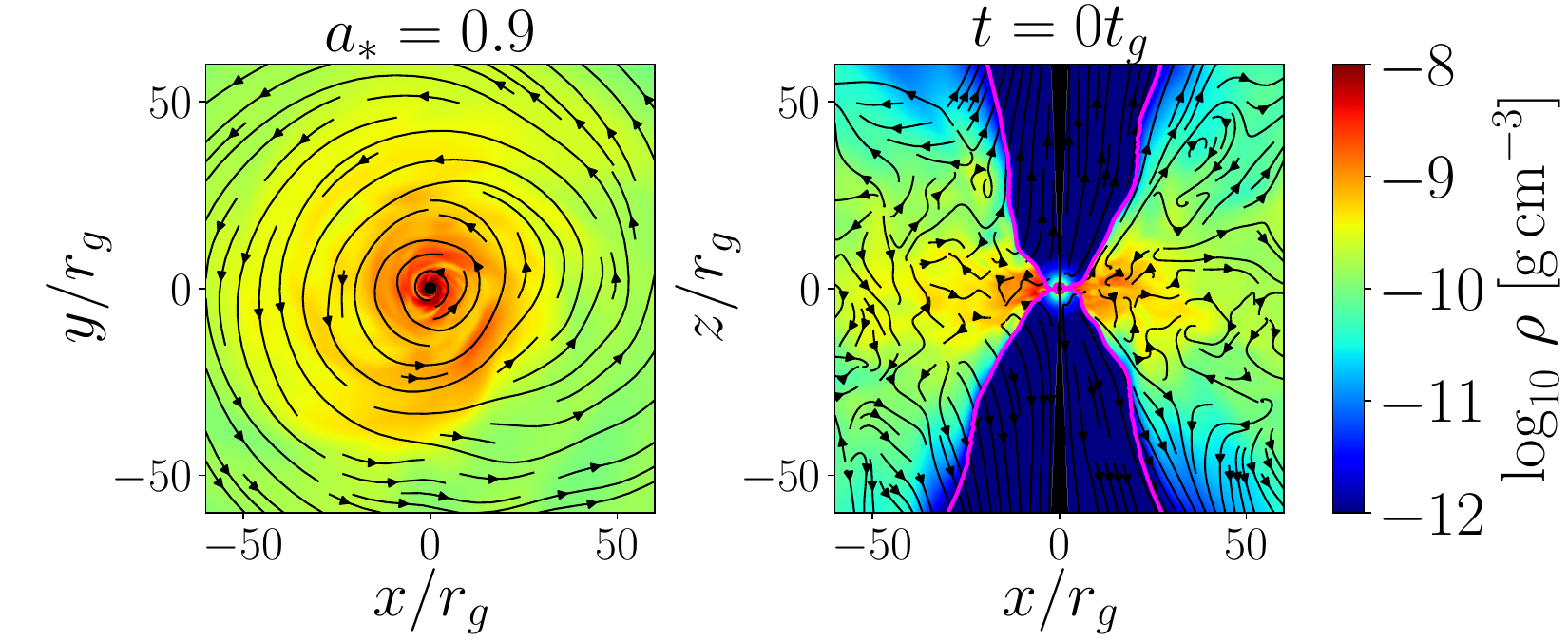}
    \caption{Initial simulation state for each BH spin. We show the gas density (colors), velocity (streamlines), and jet boundary ($\sigma=1$, pink line).}
    \label{fig:initialstate}
\end{figure}

\subsection{Initial Conditions}

The MAD model of jetted TDEs presented by \cite{Tchekhovskoy2014} assumes a powerful magnetic flux threaded the BH and disk from the time of detection until the jet shutoff. Such a system has yet to be realized in global TDE simulations with realistic BH mass and stream properties due to the substantial resources needed. In order to study a strongly magnetized disk which can be applied to TDEs, we first initialize the simulation domain with a magnetized torus and evolve it until the accretion flow becomes MAD. We then inject a stream into the simulation domain as in \citet{Curd2021}, allow the stream and disk to interact, and study how the presence of a stream changes the dynamics compared to a typical MAD system. We note that our methods are similar to that of \citet{Chan2019}, but they study systems where the disk and stream are misaligned initially and the disk is geometrically thin. Here we inject the stream along the disk midplane. 

We start with a torus of gas in hydrostatic equilibrium threaded by a large-scale poloidal magnetic field and its angular momentum aligned with the BH spin axis (or $z$-axis). This initial gas serves as a reservoir for magnetic flux. Since the evolution in our simulations is driven by $f_\rho$, we do not expect major differences in a more realistic setup. 

Most thermal TDEs dissipate $\sim10^{51}$ erg by the time they reach their peak brightness (e.g. \citealt{Mockler2021}). Assuming this radiation is of a disk origin, this places rough constraints on the amount of bound mass in the disk since $E_{\rm{bol}}\approx0.1M_dc^2$, assuming a radiative efficiency of 10\%.\footnote{We note that 10\% radiative efficiency is typically assumed in accretion models. However, TDEs could be highly radiatively inefficient. \cite{Curd2022} find efficiencies as low as $\sim$1\%, which would imply a higher maximum allowable disk mass in our initial conditions.} We do not differentiate between `disk' and `debris cloud' once the system is initialized, but the initial mass in the simulation is only $\approx9.6\times10^{-3}M_\odot$ with units scaled for a BH mass of $10^6\, M_\odot$, which guarantees that whatever mass winds up in a disk does not violate the maximum of $\sim0.01M_\odot$ based on the energy dissipation limits in TDEs.

This argument assumes a disk origin of observed radiation, but early emission may be primarily from shocks and/or an extended, Eddington limited envelope. It is also worth noting that even in deeply penetrating TDEs where kinetic dissipation due to stream self-intersection is substantially higher  \citep[e.g.][]{Liptai2019} and the disk mass might build up faster, the escaping radiation remains near Eddington \citep{Curd2021,Huang2024}. Most of the dissipated energy goes into heating the gas and may ultimately never be observed except in the outflows it may drive. If shocks or a cooling envelope dominate the optical to X-ray bands in TDEs, the connection between total radiated energy and disk mass is ultimately invalid. With regards to our models, the key point is not the disk mass since we solve the fluid evolution in GRMHD. Instead, the key connection to TDEs is $f_\rho$, which all TDEs are expected to evolve through as the stream feeds more mass into the surrounding debris regardless of total disk mass. It is also worth noting that thermal TDEs can not be observed in the FUV, but theoretical models \citep{Dai2018,Curd2019,Thomsen2022} predict most of the emission is actually emitted in these bands, which opposes arguments for disk mass limits from the observed total energy.

We note that we did not start with $\sim0.1 M_\odot$ in the simulation domain, which is the expected debris mass near peak. Had we used a torus extending to $\sim 1000 r_g$, as expected in the early debris, we would have started with $\sim0.1 M_\odot$ since the amount of enclosed mass would scale as $M(r)\propto r^2$ for a $\rho \propto r^{-1}$ density profile. Since we only inject gas at $250 r_g$, it is unnecessary to extend the torus that far, hence the smaller initial mass. Again, the key factor is $f_\rho$ that the stream encounters, not the total mass in the surrounding medium. Since the stream cross section is small (on the order of $H^2$), the interaction between the stream and debris becomes stronger as the density increases (hence smaller radii are more important dynamically).

The peak magnetic flux in the initial torus is $\approx8\times10^{30} \rm{G \ cm^{2}}$, well over the minimum required to thread the BH horizon in \textit{Swift} J1644+57 at the time of detection of $\sim10^{30} \rm{G \ cm^{2}}$ \citep{Tchekhovskoy2014}. Note that this is a necessary choice in our initial conditions to achieve an initial MAD state before introducing the stream. Such a high magnetic flux cannot originate from main sequence stars, which have a surface magnetic flux ranging from $10^{23-25} \rm{G \ cm^{2}}$. However, magnetic flux dragged in from a fossil disk by the stream or a dynamo which converts the mostly toroidal stream field into poloidal flux can amplify the magnetic flux at the BH above the MAD threshold (\citealt{Kelley2014,Liska2020}, see Appendix \ref{sec:appA} for a brief review).

From the torus initial conditions, the magnetorotational instability naturally develops and drives accretion onto the BH, which ultimately drags in magnetic field which saturates at the MAD state. We perform two such initial simulations (one for each BH spin) and evolve this initial stage for $15,000 t_g$, which is long enough for the magnetic field to saturate. We give additional details of the initial torus and time evolution of our initial setup in Appendix \ref{sec:appC}. The simulation state for each BH spin after the initial evolution before stream injection is shown in \autoref{fig:initialstate}.

\subsection{Stream Injection}

The BH mass is set to $10^6M_\odot$, though this only sets the units since GRMHD is scale free. To inject the stream, we assume orbital properties applicable to the disruption of a $1M_\odot$ star on a parabolic trajectory (eccentricity $e=1$) around a $10^6M_\odot$ BH and follow the injection methodology described in \citet{Curd2021} with a few modifications. We reproduce relevant expressions from \citet{Curd2021} below for completeness.

We describe the disruption in terms of the impact parameter, $\beta$, which is defined as the ratio between the tidal radius and pericenter separation such that $\beta \equiv R_t/R_p$. We choose $\beta=4$ for BH spin $a_*=0$ models and $\beta=7$ for $a_*=0.9$. The self-intersection radius (ignoring interaction between the stream and disk) is $\sim50\,r_g$ for all models. Note that we choose $\beta>1$ for numerical reasons. The self-intersection radius is roughly inversely proportional to $\beta$, and simulating a $\beta=1$ case for $10^6\,M_\odot$ BH would require that we have high azimuthal resolution at $r\sim 10^3r_g$, which is not numerically feasible without adaptive mesh refinement and/or GPU acceleration.

We apply the 'frozen in' approximation to estimate the spread in binding energy \citet{Stone2013}:
\begin{equation} \label{eq:bindingenergy}
  \Delta\epsilon \approx 4.3\times10^{-4} \dfrac{m_6^{1/3}m_*^{2/3}}{r_*}c^2.
\end{equation} 
We set the binding energy of the stream to that of the most bound component, $\epsilon_{\rm{inj}}=\epsilon_{\rm{mb}} = \epsilon_* - \Delta\epsilon/2$. Here $\epsilon_*$ is the initial orbital binding energy of the star, which is zero since we assume a parabolic orbit. We note that this is not accurate for late times in a TDE and $\epsilon$ of incoming material will slowly approach zero, but we maintain this assumed binding energy for all simulations for simplicity. The orbit of the disrupted star is assumed to be aligned with the equatorial plane of the BH spin vector.

For each simulation we fix $\dot{M}_{\rm{inj}}$ (and correspondingly $\rho_{\rm{inj}}$) to be constant since the simulation time is much shorter than the fallback time. To compare our model calculations with the expected $f_\rho$ curves in \autoref{sec:densitycontrast}, we measure the corresponding density contrast $f_{\rho,0}$ at $r=250r_g$ (see \autoref{tab:table1}). We set the gas temperature $T_{\rm{inj}}=10^5$ K, gas pressure $p_{\rm{inj}}=k_BT_{\rm{inj}}/\mu_{\rm{gas}} m_p$ \footnote{Here $k_B$ is the Boltzmann constant, $m_p$ is the mass of a proton, and $\mu_{\rm{gas}}$ is the mean molecular weight assuming Solar metallicity.}, and injection radius $R_{\rm{inj}}=250\,r_g$. Due to resolution limitations, we assume $(H/R)_{\rm{inj}}=0.05$, which subtends only 6 cells in $\vartheta$ and 2 cells in $\varphi$ in our grid. The angular momentum is fixed to the value corresponding to the pericenter radius of the TDE stream $l=\sqrt{2R_{\rm{p}}}$, from which we obtain the $\varphi$ velocity $v^\varphi=l/R_{\rm{inj}}$. The total velocity is then set by
\begin{equation}
    v_{\rm{inj}} = \sqrt{\dfrac{2}{R_{\rm{inj}}} + 2\epsilon_{\rm{inj}}} ~,
\end{equation} 
from which we obtain the radial velocity, $v^r=-\sqrt{(v_{\rm{inj}})^2-(v^\varphi)^2}$. We inject a weak toroidal magnetic field with the stream by setting
\begin{equation}
    B^r_{\rm{inj}}=\dfrac{p_{\rm{inj}}\beta_{g,\rm{inj}}}{\sqrt{g^{rr}}}\cos\left(\dfrac{|\vartheta-\pi/2|}{(H/R)_{\rm{inj}}}\pi\right),
\end{equation}
where $\beta_{g,\rm{inj}}=10^{-3}$ is the ratio magnetic and gas pressure in the injection cells. The other field components are set to $B^\vartheta_{\rm{inj}}=B^\varphi_{\rm{inj}}=0$.

\begin{table}
    \centering
    \begin{tabular}{ l c c c c c c}
        \hline
        \hline 
        Model & $a_*$ & $\beta$ & $\dot{M}_{\rm{inj}}$ & $f_{\rho,0}$ & $t_{\rm{start}}$ & $t_{\rm{end}}$  \\
                & & & $(\dot{M}_{\rm{Edd}})$ & & ($10^4t_g$) & ($10^4t_g$) \\
        \hline
        \texttt{m00f0.3b4} & 0 & 4 & 1 & $0.3$ & 0 & 2 \\
        \texttt{m00f0.003b4} & 0 & 4 & 100 & $0.003$ & 0 & 2 \\
        \texttt{m09f1b7A} & 0.9 & 7 & 1 & $1$ & 0 & 2 \\
        \texttt{m09f0.1b7A} & 0.9 & 7 & 10 & $0.1$ & 0 & 3.5 \\
        \texttt{m09f0.01b7} & 0.9 & 7 & 100 & $0.01$ & 0 & 3.5 \\
        \texttt{m09f1b7B} & 0.9 & 7 & 1 & $1$ & 2 & 3.5 \\
        \texttt{m09f0.1b7B} & 0.9 & 7 & 10 & $0.1$ & 2 & 7 \\
    \hline
    \end{tabular}
    \caption{Here we describe the relevant parameters of each model presented in this work. Models \texttt{m09f1b7B} and \texttt{m09f0.1b7B} are restarts of \texttt{m09f0.01b7} from 20,000 $t_g$ with the injection rate lowered to increase the initial density contrast $f_{\rho,0}$ (measured at $r=250r_g$) to study how an evolved system changes once self-intersection is weakened. We give the total stream mass introduced during the stream/disk interaction phase in Table \ref{tab:tableC1}.}
    \label{tab:table1}
\end{table}

\subsection{Connection to Astrophysical TDEs}

As we do not perform global disk formation simulations, what is the origin of the initial MAD disk in each of our simulations? While the initial MAD is necessary to provide enough magnetic flux, what is its connection to real TDEs? At all $f_\rho$, the initial MAD is assumed to represent the distribution of gas which resulted from TDE stream material which already returned to pericenter at an earlier time than when we start our simulation. We assume that the debris became MAD to satisfy the conditions asserted to be present in jetted TDEs by \cite{Tchekhovskoy2014}, who assume a MAD state at all points in jetted TDEs prior to the X-ray shut-off, hence the gas density distribution described in \autoref{sec:densitycontrast} is a reasonable approximation of the TDE debris.

At $f_\rho \ll 1$, where the stream will bore through the initial gas easily, the initial MAD is interpreted as some initial debris formed from the TDE stream. For instance, for our $f_\rho=0.01$ model, the initial MAD represents debris which resulted from the first $\sim60$ days of post-peak evolution for a $m_6=1$ BH (see \autoref{fig:frhoex}). The debris carries magnetic flux which we assume is high enough to create a MAD. Realistically this field may have been obtained from a fossil disk or a dynamo process. Note that neither of these mechanisms are captured during our simulation. Our initial conditions assume that the disk is circularized and is small in scale (only extending to $\sim 300 r_g$), but TDEs may lead to larger scale disks which are not well circularized initially. In fact, as we show later in the text, the stream causes the debris to become highly eccentric. Nevertheless, the disk in TDEs which are MAD (i.e. jetted TDEs such as \textit{Swift} J1644+57) are expected to have a density profile similar to $\rho \propto r^{-1}$, and this is ultimately what is driving the dynamics in our simulations.

At $f_\rho \gtrsim 0.01$, the stream is expected to merge with the initial debris and circularization efficiency would be higher. The initial MAD would be interpreted as the late stage, circularized debris formed from the TDE stream. For example, for our $f_\rho=0.1$ model, the initial MAD represents debris which resulted from the first $\sim300$ days of post-peak evolution for a $m_6=1$ BH (see \autoref{fig:frhoex}). The assumed origin of the magnetic flux would be the same as above.

In our setup, the stream perturbs the system until a new quasi-equilibrium is achieved. This quasi-equilibrium is $f_\rho$ dependent and we assert that this is a reasonable approximation of a MAD, jetted TDE system with the same $f_\rho$, density profile, and similar orbital properties.

The debris cloud/disk geometry in real TDEs may result in a different density profile, which could potentially place more mass at larger radii. This would lead to an earlier transition to weak self-intersection. Nevertheless, the general time evolution of $f_\rho$ would still be similar, with $f_\rho$ increasing over time.  

Our simulations apply primarily to TDEs with $R_p\lesssim 10 r_g$, which could be either $\beta=1$ TDEs around more massive BHs ($M_{\rm BH}\gtrsim 10^7 M_\odot$) or more penetrating TDEs around lower mass BHs, with the former being more probable. While higher mass BHs would lead to more bound gas than used in our models (\autoref{eq:bindingenergy}), the scaling of $m_6^{1/3}$ only changes the binding energy by a factor of $\sim 1 -$ a few, so the braking effect from the stream-disk interaction seen in our models should not depend strongly on BH mass. While the larger self-intersection radius in $\beta=1$ TDEs in the most typically assumed $M_{\rm BH}=10^6 M_\odot$ case will cause weaker outflows, the asymmetrical flow could cause similar tilt perturbations to the jet and disk.

\section{Definitions} \label{sec:definitions}

In this section, we discuss the units adopted throughout the text and provide brief descriptions of quantities used to study the \textsc{KORAL} simulation data. 

Throughout this work, we use gravitational units to describe physical parameters. For distance we use the gravitational radius $r_g\equiv GM_{\rm{BH}}/c^2$ and for time we use the gravitational time $t_g\equiv GM_{\rm{BH}}/c^3$, where $M_{\rm{BH}}$ is the mass of the BH. Often, we set $G=c=1$, so the above relations would be equivalent to $r_g=t_g=M_{\rm BH}$. Occasionally, we restore $G$ and $c$ when we feel it helps to keep track of physical units.

We adopt the following definition for the Eddington mass accretion rate:
\begin{equation} \label{eq:mdotEdd}
  \dot{M}_{\rm{Edd}} = \dfrac{L_{\rm{Edd}}}{\eta_{\rm NT} c^2},
\end{equation}
where $L_{\rm{Edd}} = 1.25\times 10^{38}\, (M_{\rm{BH}}/M_\odot)\, {\rm erg\,s^{-1}}$ is the Eddington luminosity, $\eta_{\rm{NT}}$ is the radiative efficiency of a thin disk around a BH with spin parameter $a_*$ (which is often referred to as the Novikov-Thorne efficiency):
\begin{equation} \label{eq:etaNT}
  \eta_{\rm{NT}} = 1 - \sqrt{1 - \dfrac{2}{3 r_{\rm{ISCO}}}},
\end{equation}
and $r_{\rm{ISCO}}=3+Z_2 - \sqrt{(3-Z_1)(3+Z_1+2Z_2)}$ is the radius of the Innermost Stable Circular Orbit (ISCO, \citealt{1973blho.conf..343N}) in the Kerr metric, where $Z_1 = 1 + (1-a_*^2)^{1/3}\left((1+a_*)^{1/3}+(1-a_*)^{1/3}\right)$ and $Z_2 = \sqrt{3a_*^2 + Z_1^2}$. For $a_* =$ 0 and 0.9, the efficiency is $\eta_{\rm{NT}}=$ 5.72\% and 15.58\%.

We compute the net mass inflow rate as
\begin{equation} \label{eq:mdotin}
  \dot{M}(r) = -\int_0^{2\pi}\int_0^\pi\sqrt{-g}\rho \,u^r d\vartheta d\varphi.
\end{equation}

The magnetic flux is computed as
\begin{equation} \label{eq:magflux}
  \Phi(r) = -\dfrac{1}{2}\int_0^{2\pi}\int_0^{\pi}\sqrt{-g}|B^r(r)|d\vartheta d\varphi,
\end{equation}
where $B^r$ is the radial component of the magnetic field. 

The total energy flux (excluding the rest mass flux) is computed as
\begin{equation} \label{eq:lMHD}
  L(r) = -\int_{0}^{2\pi}\int_{0}^{\pi}\sqrt{-g} (T^r_{\ \, t} + \rho u^r)  d\vartheta d\varphi.
\end{equation}

We track the time evolution of the mass accretion rate, magnetic flux, and jet power through unitless quantities evaluated at the BH horizon. We track the accretion of mass onto the BH in each simulation in Eddington units
\begin{equation} \label{eq:mdotin}
  \dot{m} = \dfrac{\dot{M}(r_H)}{\dot{M}_{\rm{Edd}}}.
\end{equation}
We quantify the magnetic field strength at the BH horizon through the normalized magnetic flux parameter \citep{2011MNRAS.418L..79T}
\begin{equation} \label{eq:eq20}
  \phi = \dfrac{\Phi(r_H)}{\sqrt{\dot{M}(r_H)}}.
\end{equation} 
For geometrically thick disks the MAD state is achieved once $\phi\sim 40-50$ \citep[see e.g.][]{2011MNRAS.418L..79T,2012JPhCS.372a2040T}.
Since the majority of the escaping energy leaves the system through the jet in MAD disks, we quantify the jet power via the total efficiency at the BH horizon
\begin{equation} \label{eq:mdotin}
  \eta=\dfrac{L(r_H)}{\dot{M}(r_H)c^2}.
\end{equation}

To determine the driving factor for angular momentum transport, we measure the effective viscosity
\begin{equation} \label{eq:alphaeff}
  \alpha_{\rm eff}=\dfrac{u^ru^{\varphi}}{c_s^2},
\end{equation}
Reynolds viscosity
\begin{equation} \label{eq:alphaRey}
  \alpha_{\rm Rey}=\dfrac{\widehat{T}_{\rm Rey}^{\hat{r}\hat{\varphi}}}{p_b + p_g},
\end{equation}
and Maxwell viscosity
\begin{equation} \label{eq:alphaMax}
  \alpha_{\rm Max}=\dfrac{\widehat{T}_{\rm Max}^{\hat{r}\hat{\varphi}}}{p_b + p_g}.
\end{equation}
Here ${\widehat{T}^{\hat{r}\hat{\varphi}}}$ is the average orthonormal $r,\ \varphi$ component of the stress-energy tensor measured in the fluid frame, $c_s$ is the sound speed, and $p_b = b^2/2$ is the magnetic pressure. Note that we have taken advantage of the fact that the stress-energy tensor can be broken into gas (Reynolds) and magnetic (Maxwell) components. That is we write \autoref{eq:Tmunu} strictly in terms of the gas or magnetic components.

We compute the eccentricity at each grid point via
\begin{equation} \label{eq:eccentricity}
  e = \sqrt{1+2\epsilon l^2},
\end{equation} 
where $\epsilon=-(u_t+1)$ is the binding energy and  $l=u_\varphi$ is the angular momentum.

To quantify the orientation of the disk and jet (or corona/funnel), we first use the magnetization to divide the fluid into 'disk' ($\sigma<1$) and 'jet' ($\sigma\geq 1$). In simulations where there is no spin, this is not a true jet since there is no mechanism to accelerate the gas to relativistic speeds. Nevertheless, this region is likely to be low optical depth and represents where X-rays are likely to escape.

Note that we transform quantities from spherical polar to cartesian coordinates $x^i=(x,y,z)$ to describe the position and angular momentum of the fluid in the following paragraphs. The angular momentum of the BH is aligned with the $z$-axis, so 
\begin{equation} \label{eq:mdotin}
  J^i_{\rm{BH}}=(0,0,a_{\rm{BH}}M).
\end{equation}
Since this term cancels when computing the tilt and precession and is meaningless for a Schwarzschild BH, we only show it here for completeness.
We derive the angular momentum of each cell in the disk using the stress energy tensor transformed to Cartesian coordinates
\begin{equation} \label{eq:Sdisk}
    S^i=[i\,j\,k]x^jT_{\rm{Cart}}^{0k},
\end{equation}
where the brackets denote the antisymmetric Levi-Cevita tensor. We then find the shell integrated, density weighted angular momentum components 
\begin{equation} \label{eq:Jdisk}
  J^i=\dfrac{\int_{0}^{2\pi}\int_{0}^{\pi}\sqrt{-g}\, w_{\rm{disk}}(\sigma)\rho \,S^i d\vartheta d\varphi}{\int_{0}^{2\pi}\int_{0}^{\pi}\sqrt{-g} \,w_{\rm{disk}}(\sigma)\rho d\vartheta d\varphi}.
\end{equation}
In the above expression, the term
\begin{equation} \label{eq:Sdisk}
    w_{\rm{disk}}(\sigma)=
     \begin{cases}  
1 , \quad  & \sigma  <1 \\
       0 , \quad  & \sigma\geq 1 \\
     \end{cases}
\end{equation}
is used to only include the disk in integration.
We then define the tilt angle relative to the BH spin (or z-axis in the zero spin case) as a function of radius
\begin{equation} \label{eq:Tdisk}
  \mathcal{T}_{\rm{disk}}(r)=\arccos{\left[\dfrac{J^z}{\sqrt{(J^x)^2+(J^y)^2+(J^z)^2}}\right]}.
\end{equation}
We also obtain the precession angle relative to the $y$-axis
\begin{equation} \label{eq:Pdisk}
  \mathcal{P}_{\rm{disk}}(r)=\arccos{\left[\dfrac{J^y}{\sqrt{(J^x)^2+(J^y)^2}}\right]}.
\end{equation}
In aligned systems, the precession angle is not a useful quantity, but once tilt sets in it can show whether the disk and jet precess together.

For the jet, we derive a position based angle. We start by finding the $\sigma$ weighted mean position for the top and bottom jet at each radius
\begin{equation} \label{eq:xjettop}
  x_{\rm{jet, top}}^i=\dfrac{\int_{0}^{2\pi}\int_{0}^{\pi/2}\sqrt{-g}\, w_{\rm{jet}}(\sigma)\sigma\, \,x^i d\vartheta d\varphi}{\int_{0}^{2\pi}\int_{0}^{\pi/2}\sqrt{-g} \,w_{\rm{jet}}(\sigma)\sigma d\vartheta d\varphi},
\end{equation}
\begin{equation} \label{eq:xjetbot}
  x_{\rm{jet, bot}}^i=\dfrac{\int_{0}^{2\pi}\int_{\pi/2}^{\pi}\sqrt{-g}\, w_{\rm{jet}}(\sigma)\sigma\, \,x^i d\vartheta d\varphi}{\int_{0}^{2\pi}\int_{\pi/2}^{\pi}\sqrt{-g} \,w_{\rm{jet}}(\sigma)\sigma d\vartheta d\varphi}.
\end{equation}
In both expressions, the term
\begin{equation} \label{eq:Sdisk}
    w_{\rm{jet}}(\sigma)=
     \begin{cases}  
0 , \quad  & \sigma  <1 \\
       1 , \quad  & \sigma\geq 1 \\
     \end{cases}
\end{equation}
is used to explicitly exclude the disk from calculations.
We then calculate a tilt and precession angle based on the mean position. For example, the top jet's tilt and precession are calculated as
\begin{equation} \label{eq:Tjet}
  \mathcal{T}_{\rm{jet,top}}(r)=\arccos{\left[\dfrac{z_{\rm{jet, top}}}{\sqrt{(x_{\rm{jet, top}})^2+(y_{\rm{jet, top}})^2+(z_{\rm{jet, top}})^2}}\right]},
\end{equation}
and
\begin{equation} \label{eq:Pjet}  
\mathcal{P}_{\rm{jet,top}}(r)=\arccos{\left[\dfrac{y_{\rm{jet, top}}}{\sqrt{(x_{\rm{jet, top}})^2+(y_{\rm{jet, top}})^2}}\right]}.
\end{equation}
The same expressions are used for the bottom jet except with the mean coordinates $x_{\rm{jet, bot}}^i$. For both the disk and jet, we report the average tilt and precession angles over $10\leq r/r_g\leq100$.

We quantify the jet opening angle by computing the solid angle it subtends in a flat spacetime:
\begin{equation} \label{eq:Omegajettop}
  \Omega_{\rm{jet, top}}(r)=\int_{0}^{2\pi}\int_{0}^{\pi/2}\, w_{\rm{jet}}(\sigma)\sin(\vartheta)\cos(\vartheta)d\vartheta d\varphi
\end{equation}
\begin{equation} \label{eq:Omegajetbot}
  \Omega_{\rm{jet, bot}}(r)=-\int_{0}^{2\pi}\int_{\pi/2}^{\pi}\, w_{\rm{jet}}(\sigma)\sin(\vartheta)\cos(\vartheta)d\vartheta d\varphi.
\end{equation}
Note the minus sign in \autoref{eq:Omegajetbot} is to account for the negative introduced by $\cos(\vartheta)$. We compute an average solid angle
\begin{equation} \label{eq:Omegajet}
  \Delta\Omega(r)=\dfrac{\Omega_{\rm{jet, top}}(r)+\Omega_{\rm{jet, bot}}(r)}{2}.
\end{equation}
We relate this to the mean jet width under the assumption of a conical geometry
\begin{equation} \label{eq:wjet}
  \mathcal{W}(r) = r\sin \biggl{(}\arccos[1-\Delta\Omega(r)/2\pi]\biggr{)}.
\end{equation}

\section{Results} \label{sec:results}

\subsection{Stream/Disk Dynamics} \label{sec:dynamics}

\begin{figure*}
    \centering{}
    \begin{subfigure}[b]{0.4\textwidth}
    \includegraphics[width=\columnwidth]{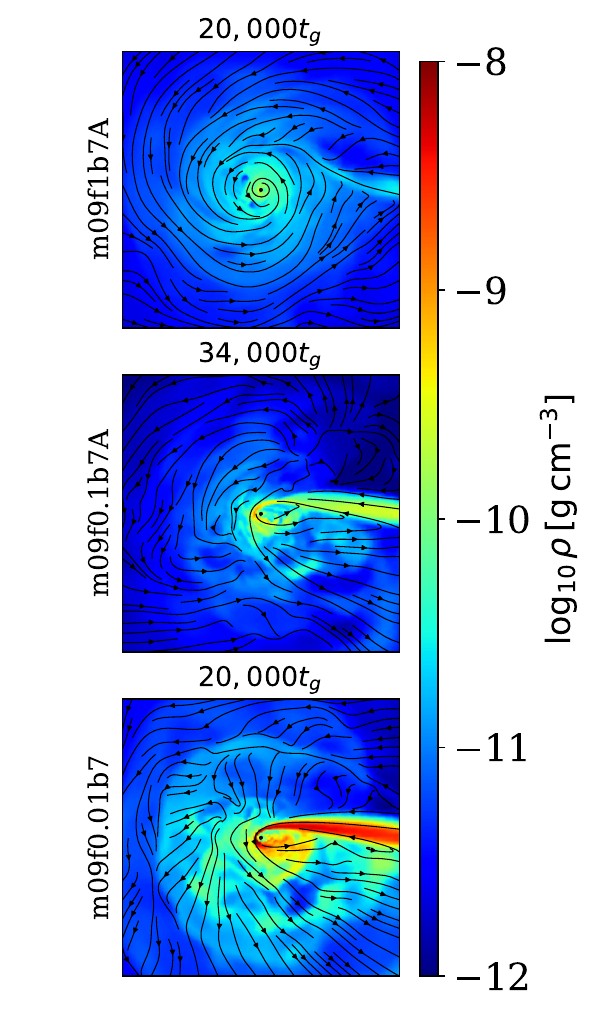}
  \end{subfigure}
    \begin{subfigure}[b]{0.4\textwidth}
    \includegraphics[width=\columnwidth]{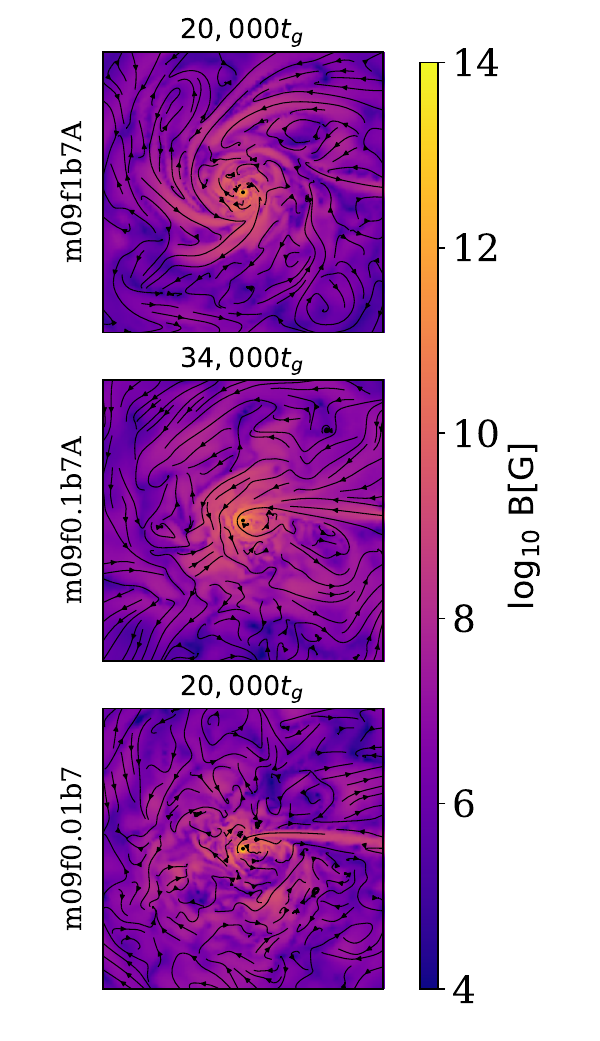}
  \end{subfigure}
    \caption{Here we show the gas density (colors, left panels), velocity field (stream lines, left panels), magnetic field strength (colors, right panels), and magnetic field (stream lines, right panels) for an equatorial slice of each of the $a_* = 0.9$ models for $f_\rho = $ 0.01 (bottom row), 0.1 (middle row), 1 (top row). Each figure spans a region of $480 r_g \times 480 r_g$ centered around the BH. We describe the figure in \autoref{sec:dynamics}.}
    \label{fig:Large_Scale_maps}
\end{figure*}

\begin{figure}
    \centering{}
	\includegraphics[width=0.7\columnwidth]{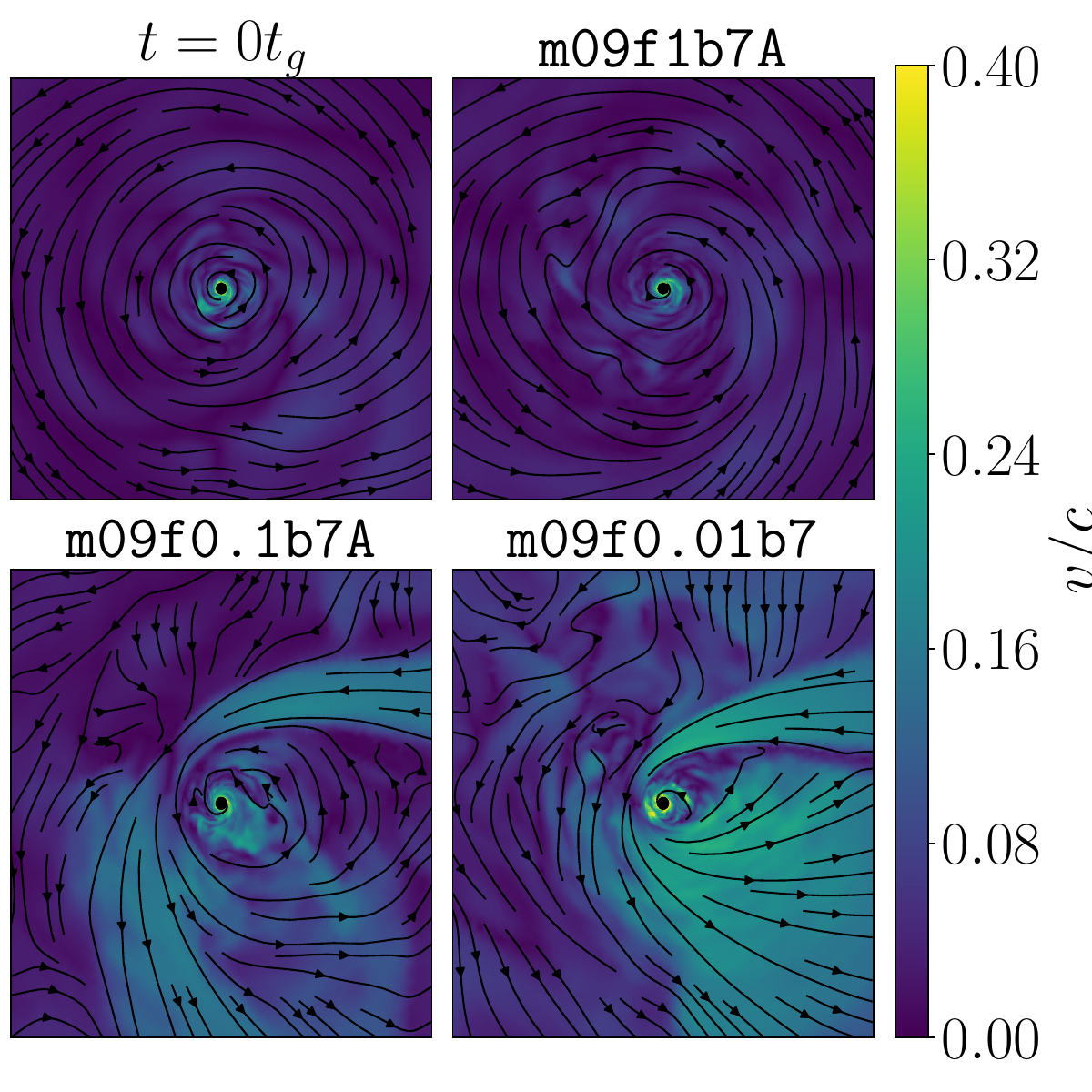}\\
    \caption{Here we show the velocity (colors) and velocity field vector (stream lines) for an equatorial slice of each of the $a_* = 0.9$ models for $f_\rho = $ 0.01 (bottom right), 0.1 (bottom left), 1 (top right). We also show the velocity field for the initial conditions on the top left for comparison. Each panel shows in $120 r_g \times 120 r_g$ region centered around the BH. See \autoref{sec:dynamics} for a description of the figures.}
    \label{fig:m09_v}
\end{figure}

\begin{figure}
    \centering{}
	\includegraphics[width=0.7\columnwidth]{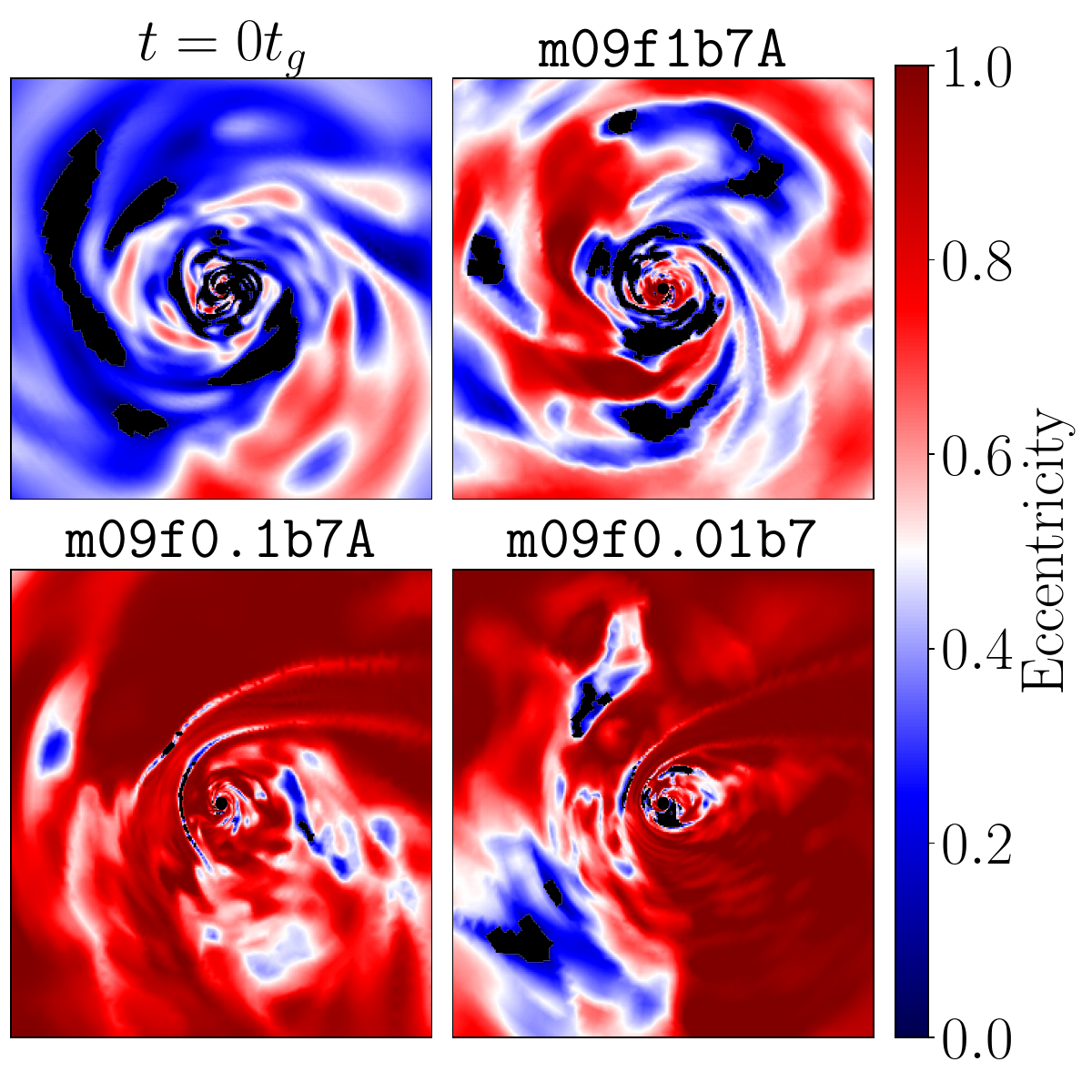}\\
    \caption{Here we show the eccentricity (colors) for an equatorial slice of each of the $a_* = 0.9$ models for $f_\rho = $ 0.01 (bottom right), 0.1 (bottom left), 1 (top right). We also show the eccentricity for the initial conditions on the top left for comparison. Each figure spans a region similar to \autoref{fig:m09_v}. See \autoref{sec:dynamics} for a description of the figures.}
    \label{fig:m09_ecc}
\end{figure}

We show the large scale structure of models with $f_\rho = 0.01, 0.1, 1$ and $a_* = 0.9$ in \autoref{fig:Large_Scale_maps} (\texttt{m09f1b7A}, \texttt{m09f0.1b7A}, \texttt{m09f0.01b7}). When $f_\rho = 0.01$, the ram pressure from the disk is negligible, and the system evolves much like disk formation simulations initialized with no initial disk \citep{Sadowski2016, Curd2021}. The stream dissipates a negligible amount of orbital energy on its way to pericenter, where it goes through a nozzle shock due to vertical compression and self-intersects at roughly the self-intersection radius (see bottom left panel in \autoref{fig:Large_Scale_maps} and bottom right panel in \autoref{fig:m09_v}). Similar to \citet{Curd2021}, the nozzle shock is poorly resolved, so we do not discuss it throughout this work. For the same BH mass with $\beta=1$, \cite{Bonnerot2022} find a minimum nozzle scale-height ($H/R$) on the order of $\sim10^{-5}$ while our numerical grid has a minimum angular extent of $\sim10^{-2}$ radians at the mid-plane, which is orders of magnitude larger. Although the nozzle is under-resolved, our resolution study in \autoref{sec:appD} suggests that enhanced resolution does not significantly impact the self-intersection and jet tilt, which are our primary focus. Assessing the impact of fully resolving the nozzle shock on our simulations is beyond the scope of this work due to computational limitations; however, it is worth noting that \cite{Bonnerot2022} find minimal impact of the nozzle shock on the subsequent self-intersection efficiency when the stream remains on the same orbital plane (as in our simulations). Bound and unbound gas is produced by the self-intersection shock, some of which falls in and makes an accretion disk while the rest flows out and interacts with the jet and outer medium. The material which forms the accretion disk maintains a high eccentricity (See bottom right panel in \autoref{fig:m09_ecc}). Despite the low magnetic field strength injected with the stream, the disk maintains a strong magnetic field due to the pre-existing field being anchored to smaller radii by inflowing material (See bottom right panel in \autoref{fig:Large_Scale_maps}). Similar to the magnetized disk formation simulations in \citet{Curd2021}, the magnetic field in material which has gone through the self-intersection shock becomes highly disordered and turbulent. However, as we discuss later, the poloidal magnetic flux inside the self-intersection radius remains trapped and the field in the inner accretion disk remains ordered. The outflowing part is launched with velocity $\sim 0.1 c$ and produces an asymmetrical ram pressure on the jet since it is quasi-spherical. This results in a force in the $-x$ direction. We describe how this effects the disk and jet evolution in \autoref{sec:lowfrho}. 

With $f_\rho = 0.1$, we observe significant slowing of the stream on its way to pericenter, but it is not completely stopped by the disk (See middle left panel in \autoref{fig:Large_Scale_maps} and bottom left panel in \autoref{fig:m09_v}). As a consequence, the pericenter radius is shifted outward radially significantly and the self-intersection has far less kinetic energy available for dissipation. No quasi-spherical outflow is produced as a result. This may in part be due to the shock weakening due to poorer resolution at larger radii. However, this result is not unreasonable since the energy and velocity of the self-intersection outflow rapidly drop off with increasing radius since the stream self-intersects at roughly the orbital velocity. We again find a highly eccentric accretion disk forms, but we note a slight decrease in eccentricity compared with the $f_\rho = 0.01$ model due to the dissipation of orbital energy as the stream interacts with the disk (See bottom left panel in \autoref{fig:m09_ecc}). Since there is no self-intersection outflow, the magnetic field in the outer accretion disk is less turbulent. We again find anchoring of poloidal magnetic field to the BH by the inflowing material. 

With $f_\rho = 1$, the ram pressure exerted on the stream by the disk is large enough to halt the stream before it reaches pericenter. Instead, the stream is observed to mix with the accretion disk (See top panel in \autoref{fig:Large_Scale_maps}). This can clearly be seen in the velocity which closely resembles the initial MAD disk (See top panels in \autoref{fig:m09_v}). Interestingly, the stream does add eccentricity to the disk as the inflowing material reaches $e > 0.7$. The field structure closely resembles a standard MAD accretion disk (e.g. bottom panel in \autoref{fig:initialstate}) since the stream has little effect on the disk. 

The dynamics for a given $f_\rho$ are similar in the $a_* = 0$ models. Videos of each simulation can be seen in our \href{https://youtube.com/playlist?list=PL6Na55ZD3RmoJl7Rjhn6gCeAE0HWYCI0b&si=R0PLe2jj0ZjvfKHS}{YouTube playlist}.


\begin{figure*}
    \centering{}
	\includegraphics[width=\textwidth]{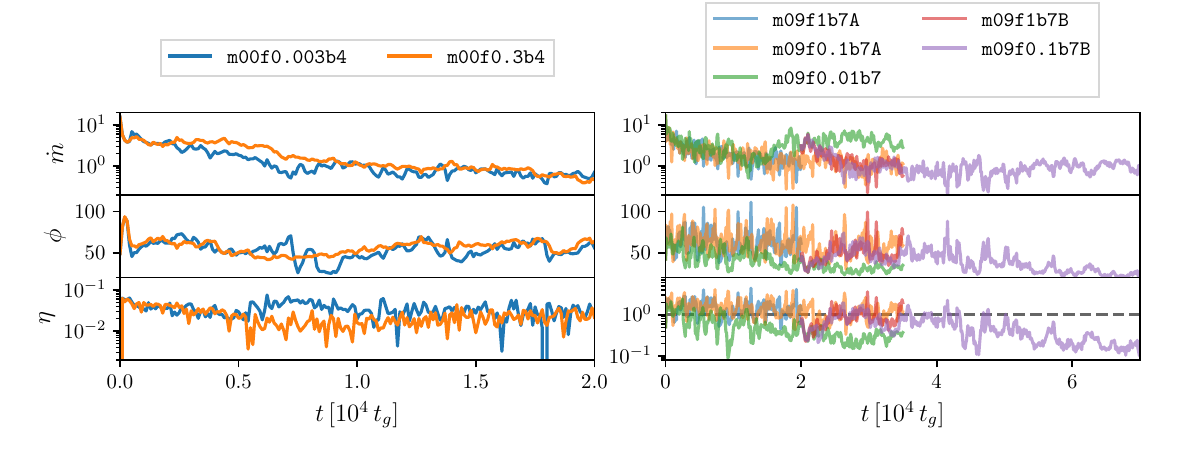}
    \caption{We show the mass accretion rate (top row), normalized magnetic flux at the BH horizon (middle row), and efficiency (bottom row) for each of the $a_* = 0$ (left column) and $a_* = 0.9$ (right column) models. Each model shows an initial decrease in the mass accretion rate as the injected stream interacts with the disk. As we discuss in \autoref{sec:scalars}, this is due to the density in the disk decreasing due to viscous spreading and mass accretion. In each model we find $\phi > 20$, which confirms that during the stream-disk interaction in TDE, disks can maintain a strong magnetic flux.} For the models where no tilt instability sets in, a MAD flux of $\phi > 50$ is maintained and a powerful jet with $\eta \approx 100-400\%$ is launched when $a_* = 0.9$. As expected, no jet is launched when $a_* = 0$ and we find similar $\eta$ for both of the $a_* = 0$ models. 
    \label{fig:scalars}
\end{figure*}


\subsection{MAD Disks Maintain Magnetic Flux and Jets During Stream-Disk Interaction} \label{sec:scalars}

We show the accretion rate, normalized magnetic flux, and efficiency at the BH horizon in \autoref{fig:scalars}. In all models save \texttt{m09f0.01b7}, the accretion rate drops from about 10 to 1 Eddington. This is due to a drop in density around the BH as the disk spreads viscously and mass is consumed by the BH. Surprisingly, there is little difference in accretion history as we vary $f_\rho$ except in \texttt{m09f0.01b7} which shows elevated accretion once the disk tilts, an effect we describe in the next section.

In all models, a MAD or semi-MAD ($\phi\sim20-50$) state is maintained. Despite the high eccentricity, magnetic field is successfully contained and does not rapidly diffuse from the horizon. This is a genuinely new result and is a bit of a surprise since \citet{Curd2021} found negligible poloidal flux accumulation when the field comes from the stream even with a favorable field configuration. Our results indicate that once poloidal flux reaches the BH, regardless of how it was obtained (i.e. fossil disk or a dynamo effect), the chaotic and eccentric disk can anchor it to the BH. We note that while \texttt{m09f0.01b7} showed a decrease in normalized magnetic flux, the total magnetic flux given by \autoref{eq:magflux} remains roughly the same. The decrease in normalized magnetic flux is due to additional accretion driven by strong shocks once the tilt sets in. See discussion in \autoref{sec:lowfrho}. 

We treat the efficiency as measured at the horizon as a proxy for the outgoing jet power. In all models with $a_* = 0.9$ we find $\eta \approx 100-400\%$ while the magnetic flux remains MAD ($\phi \gtrsim 50$). Ultimately, the jet power at larger radii may decrease especially in cases where the self-intersection outflow is strong, and the jet may interact with the disk and outflow. In addition, instabilities in the jet disk interface can lead to additional dissipation of jet power \citep{Chatterjee2019}. For models with spin $a_* = 0$, the efficiency remains much lower at $\sim 2-6\%$ since there is no jet.  

\subsection{Magnetic Stresses are Subdominant} \label{sec:alphas}
\begin{figure}
    \centering{}
	\includegraphics[width=\columnwidth]{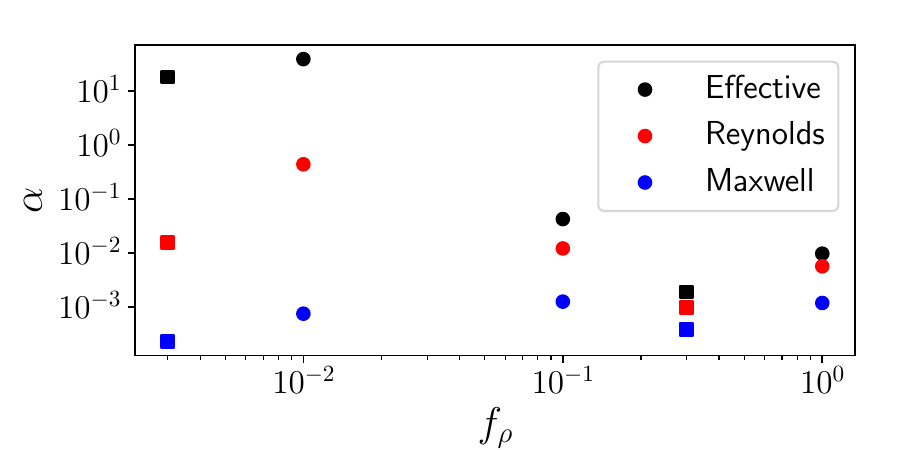}
    \caption{Here we show the radius-weighted viscosity as computed in \autoref{sec:alphas} as a function of $f_\rho$. We indicate $a_* = 0$ models as squares and $a_* = 0.9$ models as circles.}
    \label{fig:alpha}
\end{figure}
To quantify the contribution to angular momentum transport from hydrodynamic and magnetic processes, we compute a radius-weighted average of $\alpha_{\rm eff},\alpha_{\rm Rey}, $ and $ \alpha_{\rm Max}$ in the disk ($\sigma < 1$) from $r_H < r < 100 r_g$ at $t = t_{\rm end}$ \footnote{We have verified that the viscosity behaves the same across time and the qualitative properties shown in \autoref{fig:alpha} are not effected by our choice of time to perform the measurement.}. We employ radius-weighting instead of density-weighting to incorporate part of the outer disk where shocks are present into the calculation. 

We show the average viscosity in \autoref{fig:alpha} as a function of $f_\rho$. We find that the effective and Reynolds viscosity both decline as a function of $f_\rho$. Meanwhile, the Maxwell viscosity is similar across all values of $f_\rho$ with $\alpha_{\rm Max} \lesssim 10^{-3}$. At all values of $f_\rho$, the effective viscosity and the Reynolds viscosity are larger than the Maxwell viscosity. At $f_\rho \lesssim 0.01$, the effective viscosity is more than an order of magnitude larger than the Reynolds viscosity which suggests shocks dominate transport at this stage of a TDE. We observe that at $f_\rho \gtrsim 0.1$, the effective and Reynolds viscosity are of roughly the same magnitude which suggests a transition to turbulent transport. 

Our findings at $f_\rho \lesssim 0.1$ are similar to  \citet{Sadowski2016} who found even after a disk formed, the Maxwell viscosity remained subdominant by at least an order of magnitude. At $f_\rho \gtrsim 1$, the viscosity resembles some of the MAD disks in \citet{McKinney2012} which also showed a  larger Reynolds viscosity than Maxwell viscosity in spite of the powerful polodial magnetic fields.

\subsection{Disk and Jet Tilt Evolution}
\subsubsection{Low Density Contrast Jetted Model: $f_\rho=0.01$}\label{sec:lowfrho}

\begin{figure}
    \centering{}
	\includegraphics[width=\columnwidth]{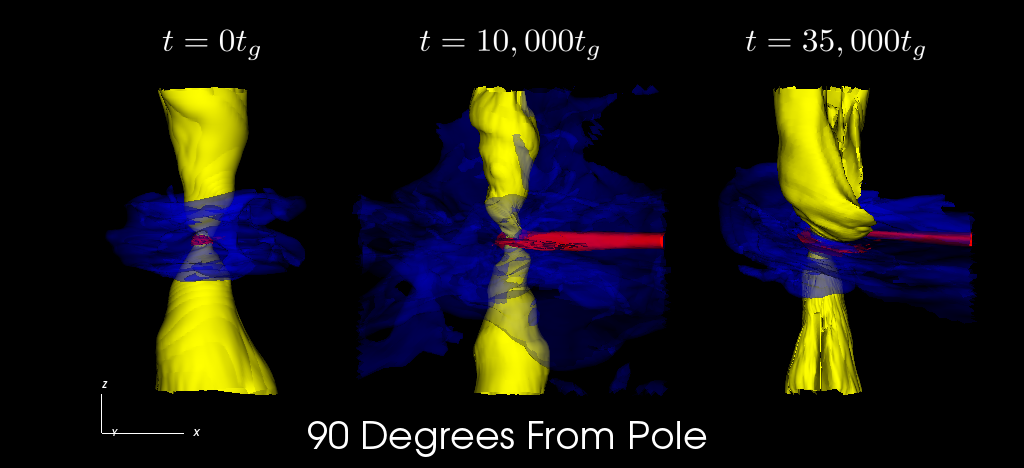}\\
	\includegraphics[width=\columnwidth]{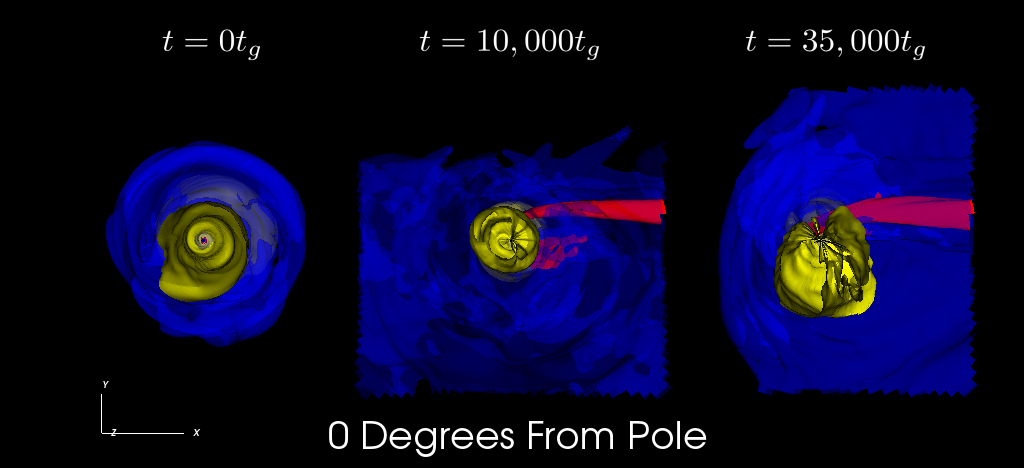}
    \caption{Volume renderings of a $200r_g \times 200r_g$ region of model \texttt{m09f0.01b7} showing the stream/disk (red), outer disk/outflow (blue), and jet (yellow) viewed edge on (top panel) and viewed down the jet axis (bottom panel). We show times in $t = 0t_g$ (left), $t = 10,000t_g$ (middle), $t = 35,000t_g$ (right). The outflow pushes on the jet laterally and begins to tilt the jet. This ultimately leads to a tilted disk and jet in the final snapshot. }
    \label{fig:volumerendering}
\end{figure}

\begin{figure}
    \centering{}
	\includegraphics[width=\columnwidth]{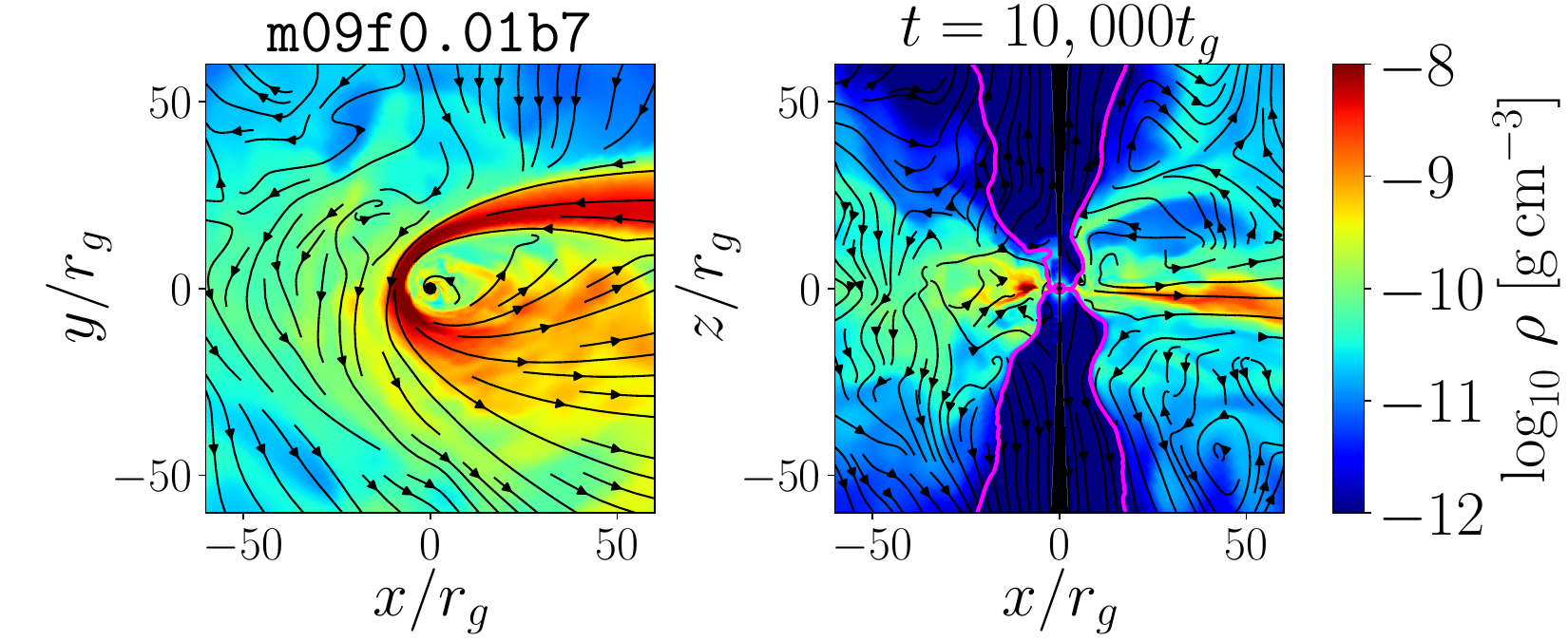}\\
	\includegraphics[width=\columnwidth]{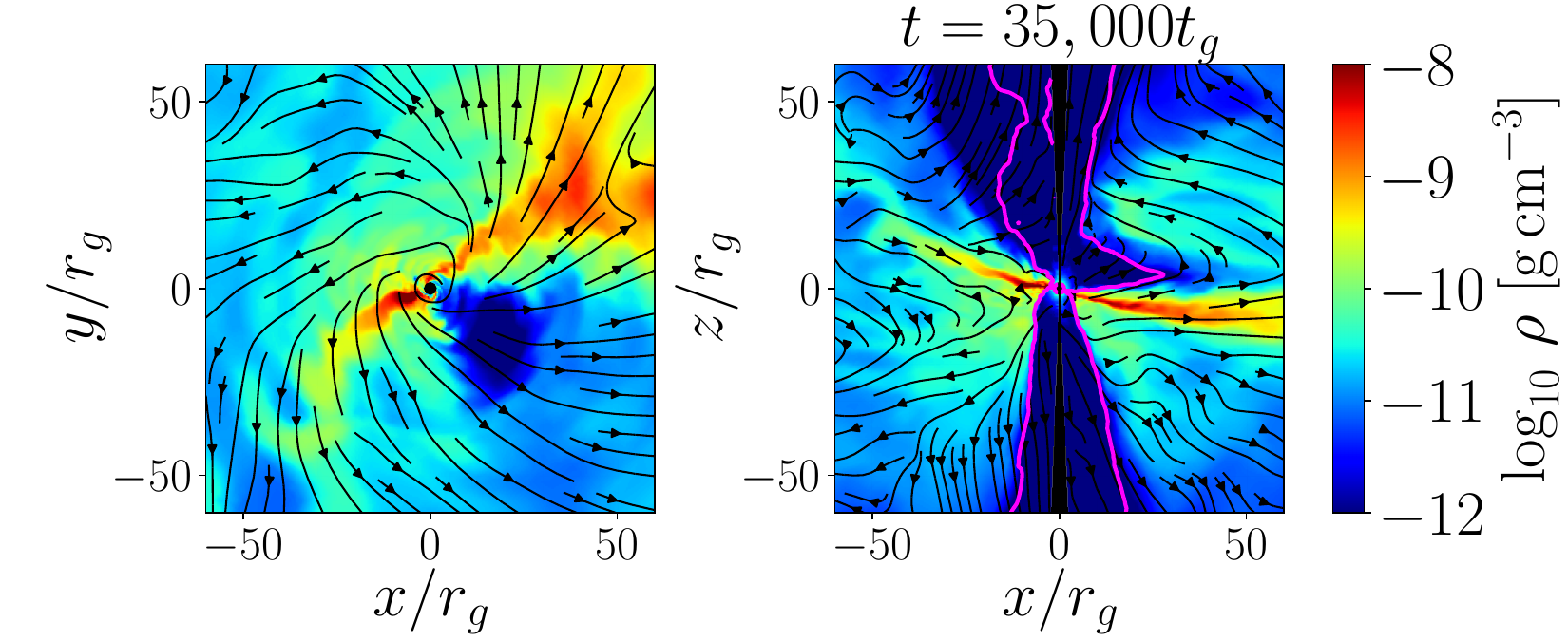}\\
    \includegraphics[width=\columnwidth]{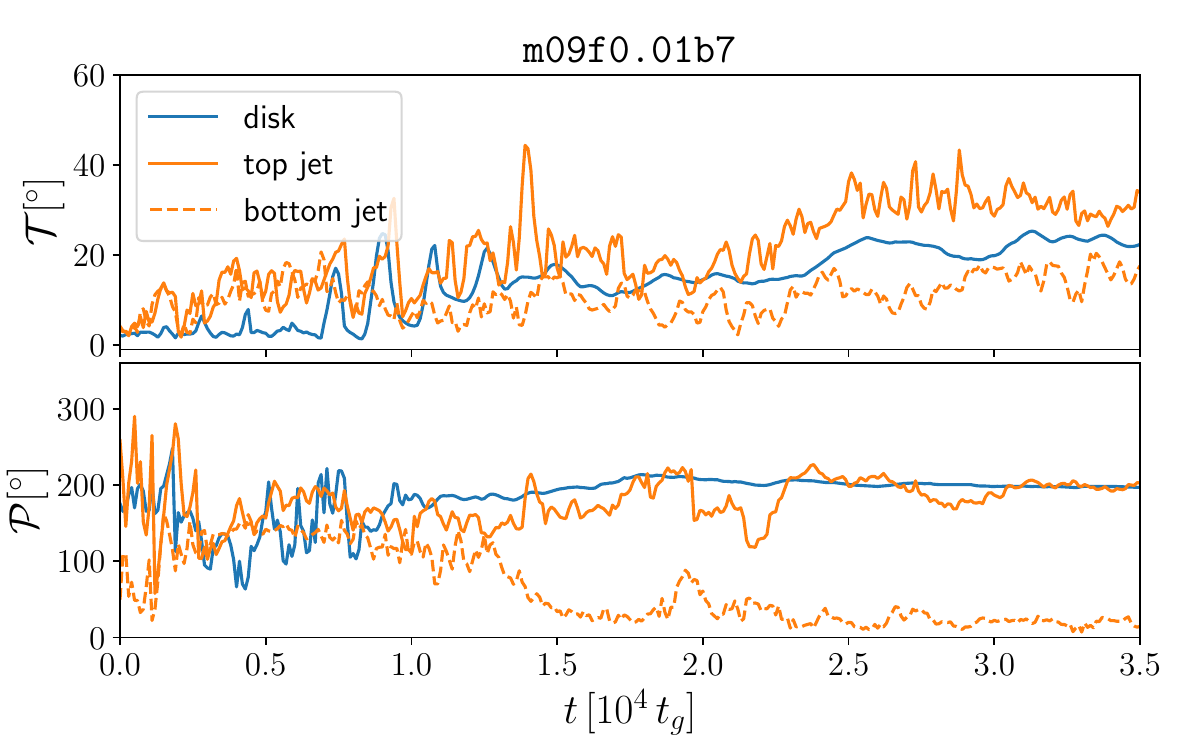}
    \caption{\textit{Top two rows:} Gas density (colors), velocity (streamlines), and jet boundary ($\sigma=1$, pink line) for \texttt{m09f0.01b7}. We show an equatorial slice (left) and vertical slice (right) spanning a region of $120 r_g \times 120 r_g$ centered on the BH. Snapshots are shown during the initial self-intersection ($t = 10^4 t_g$, first row), and at the end of the simulation after the tilt has set in (second row). \textit{Bottom two rows:} We show the tilt and precession angle for the disk and top/bottom jet over the evolution of the simulation. As the stream flows in, a quasi-spherical outflow begins to push on the jet and we see the jet tilt increase initially. At around $t = 0.6 \times 10^4 t_g$, the jet begins to perturb the disk and we observe a steady increase in the disk tilt until it roughly aligns with the jet, after which the tilt in both the disk and jet increases until they settle around a rough equilibrium state at $t = 2.5 \times 10^4 t_g$. Once the disk tilts, a feedback cycle begins due to self-intersection and magneto-spin alignment cannot realign the inner disk. The precession angle prior to the tilt setting in is not a meaningful quantity since the system is initially aligned with the BH spin. Once the system tilts, the disk and top jet share the same precession angle and we do not observe much variability in the precession. The bottom jet points in the opposite direction and is roughly $180^\circ$ out of phase with the top jet.}
    \label{fig:m09_str100Edd_b7_A}
\end{figure}

\begin{figure}
    \centering{}
	\includegraphics[width=\columnwidth]{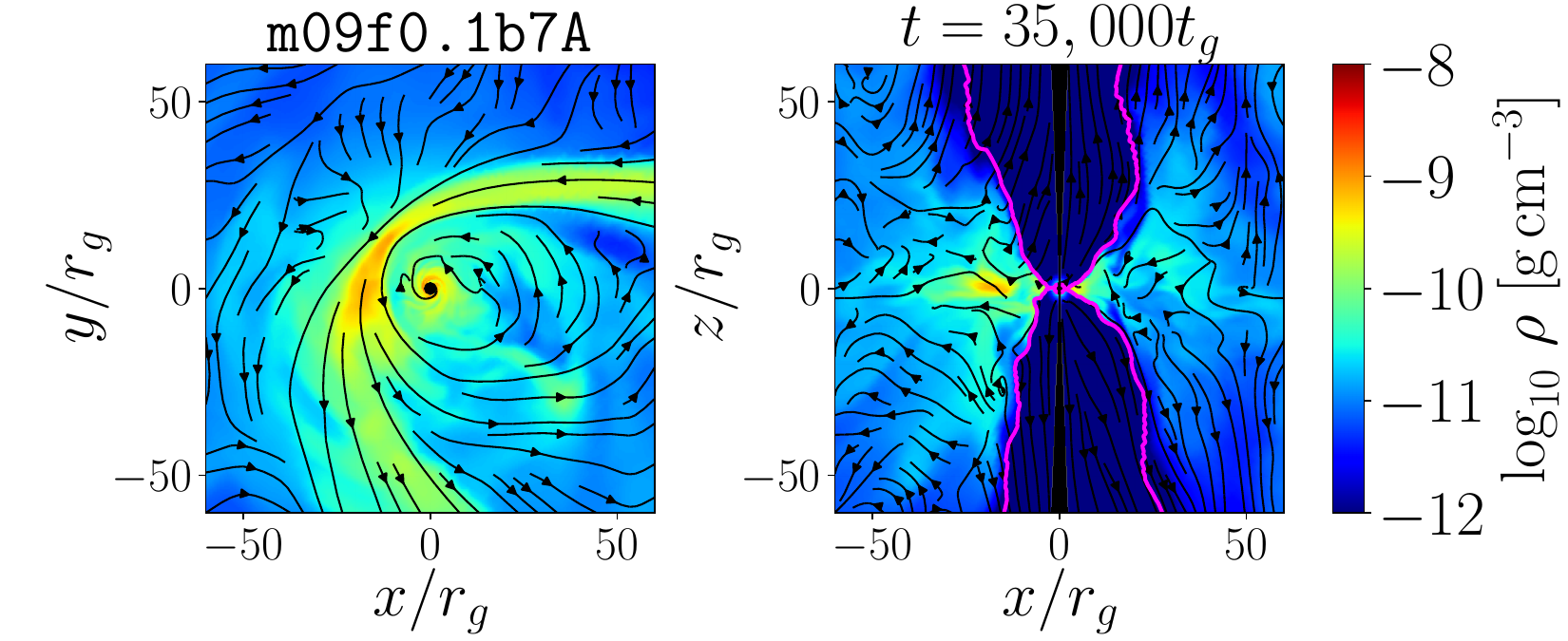}\\
    \includegraphics[width=\columnwidth]{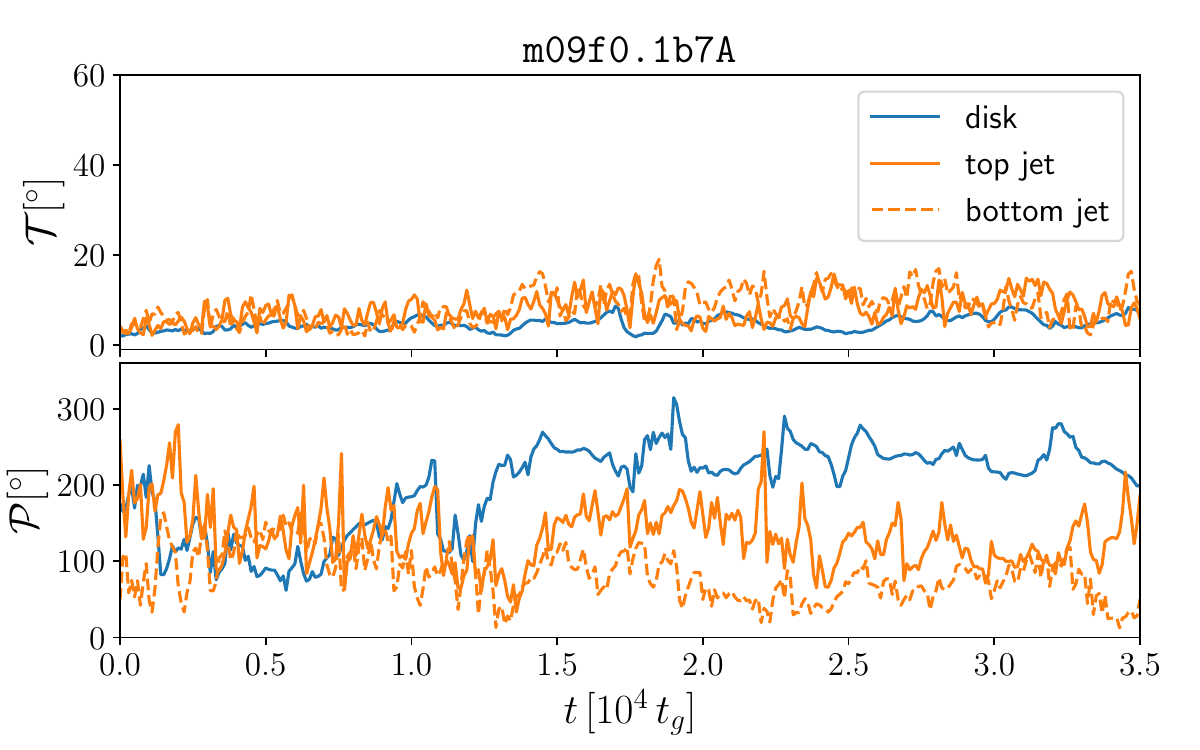}    
    \caption{\textit{Top row:} Here we show the same quantities as the top three rows in \autoref{fig:m09_str100Edd_b7_A}, but for model \texttt{m09f0.1b7A}. As we discuss in \autoref{sec:mediumfrho}, the stream loses orbital energy on its path to pericenter and the self-intersection outflow is significantly weakened which leads to a weaker perturbation on the jet. We note that the jet profile is less smooth than in the initial state (top panel in \autoref{fig:m09_str100Edd_b7_A}) due to asymmetry in the disk structure induced by the interaction with the stream. \textit{Bottom two rows:} The weak perturbation on the jet leads to a non-zero tilt measurement. However, both the disk and jet maintain low tilts with $\mathcal{T} < 10^\circ$, which confirms that strong self-intersection is needed to induce strong interaction between the jet and disk. The top and bottom jet maintain precession angles which are roughly in-phase and oscillate over time, which is typical of spin aligned MAD disks.}
    \label{fig:m09_str10Edd_b7}
\end{figure}

\begin{figure}
    \centering{}
	\includegraphics[width=\columnwidth]{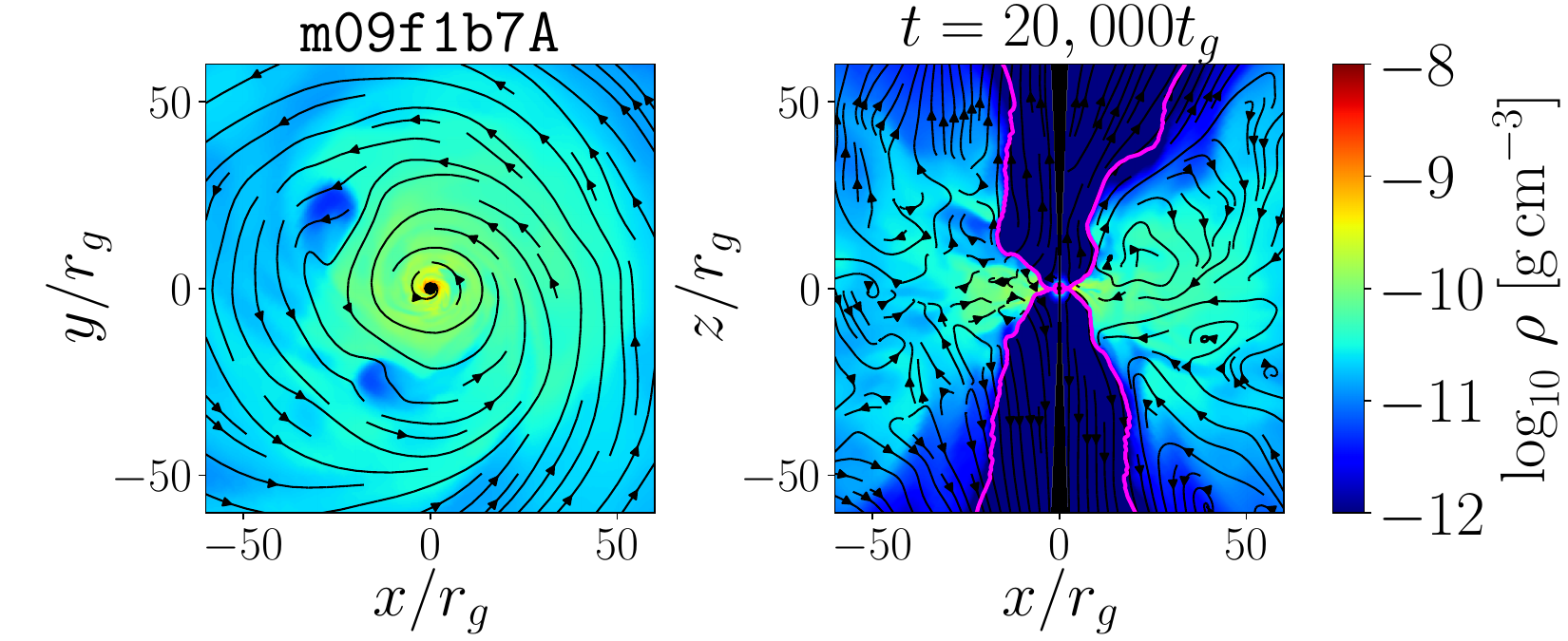}\\
    \includegraphics[width=\columnwidth]{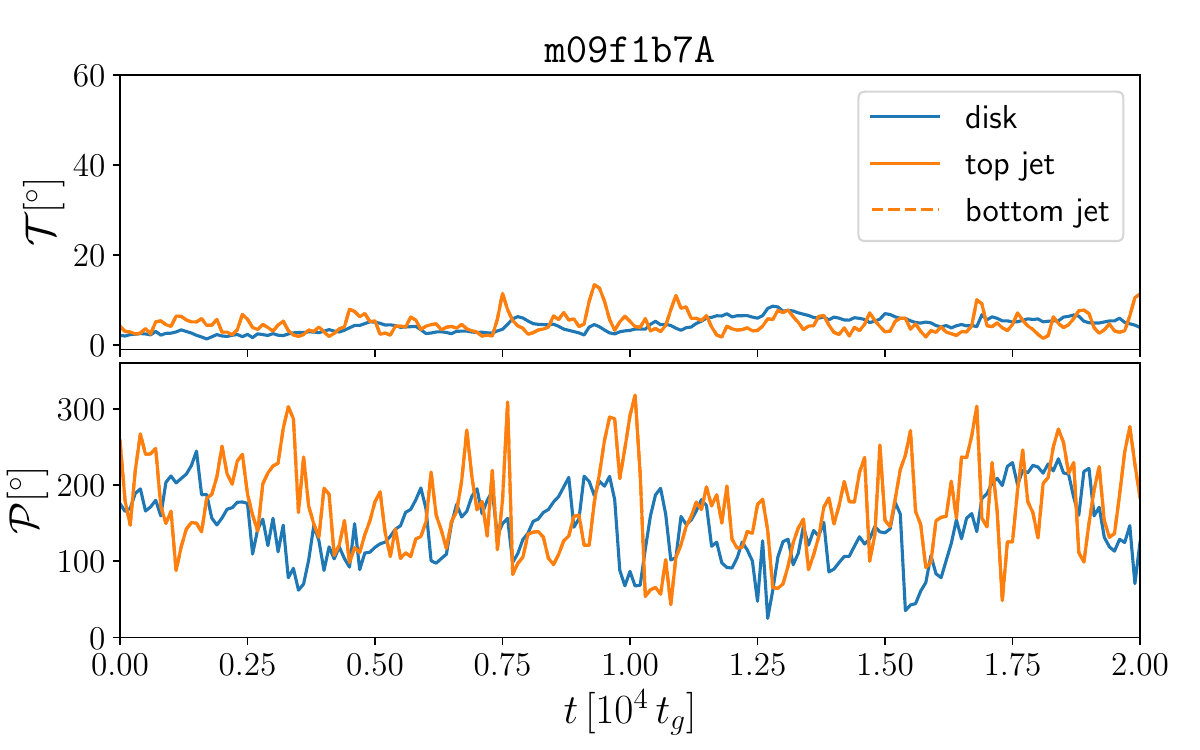}  \caption{The same as \autoref{fig:m09_str10Edd_b7}, but for model \texttt{m09f1b7A}. Since the stream is halted by the pre-existing disk, no self-intersection outflow occurs. Subsequently, the jet and disk are approximately aligned with the BH spin throughout the entire simulation. Interestingly, the added turbulence to the system during the interaction with the stream appears to perturb the jet boundary compared to the initial state. Note the precession angle of the disk is not a useful quantity since the disk is aligned with the BH.}
    \label{fig:m09_str1Edd_b7}
\end{figure}

\begin{figure}
    \centering{}
	\includegraphics[width=\columnwidth]{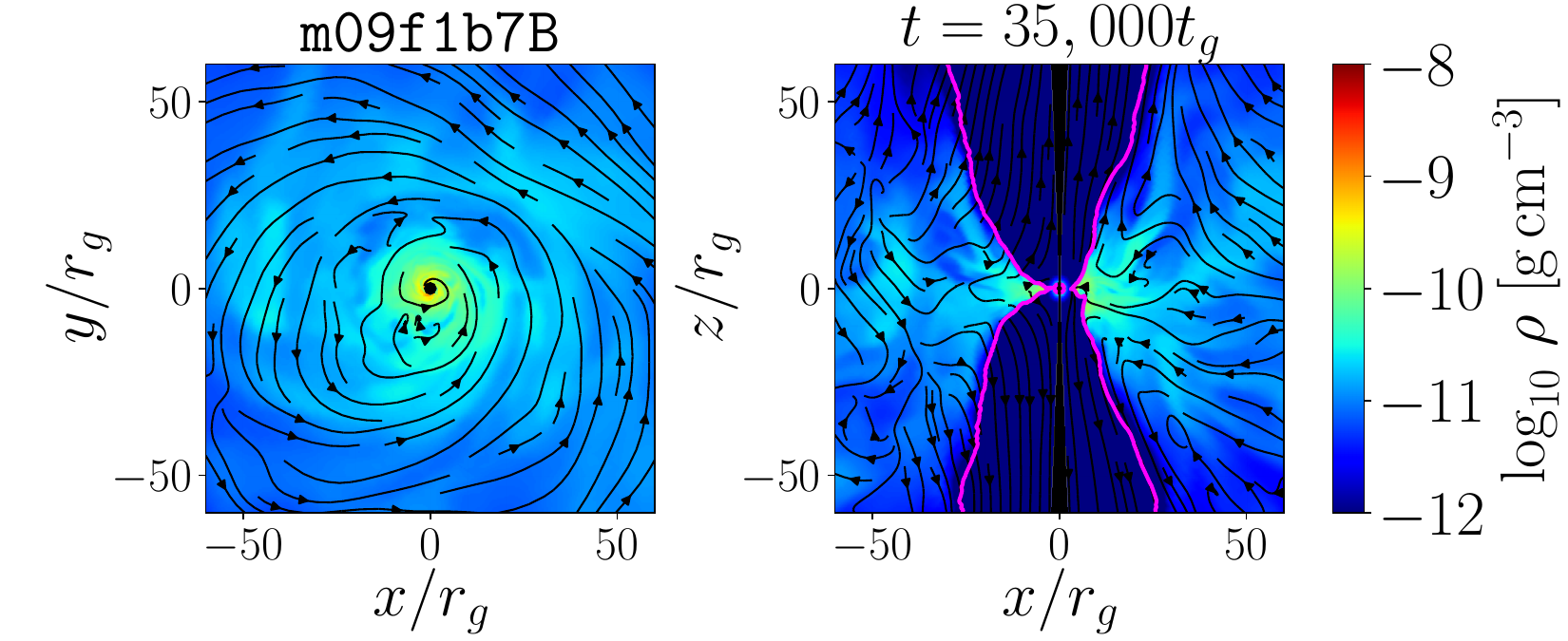}\\
    \includegraphics[width=\columnwidth]{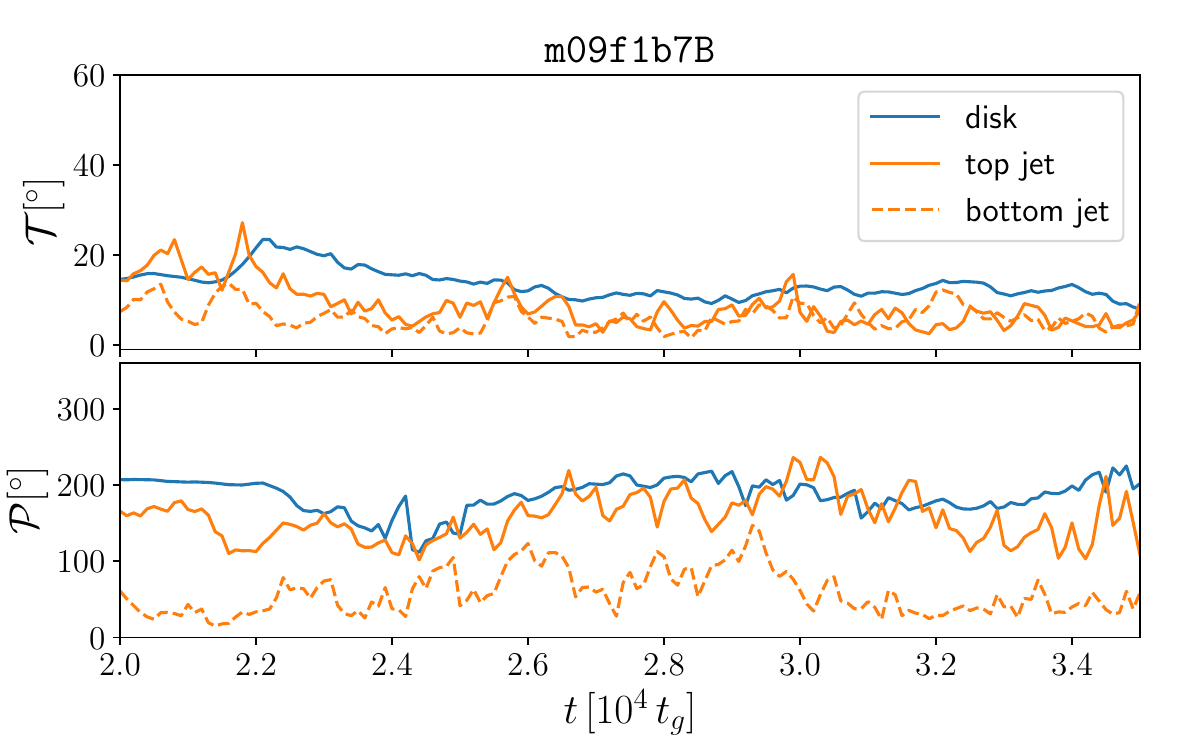}
    \caption{The same as \autoref{fig:m09_str10Edd_b7}, but for model \texttt{m09f1b7B} which is a restart of \texttt{m09f0.01b7} at $t = 2 \times 10^4 t_g$. $f_\rho$ is instantaneously increased from 0.01 to 1 at the start of the simulation. Because the stream is halted by the disk due to the change in density contrast, the self-intersection ceases shortly after we start the simulation. Without the added perturbation from a self-intersection outflow, the jet realigns with the z-axis and magneto-spin alignment rapidly realigns the disk with the BH spin. Interestingly, the top and bottom jet remain approximately $180^\circ$ out of phase even after self-intersection ceases. }
    \label{fig:m09_str100Edd_b7_B}
\end{figure}

\begin{figure}
    \centering{}
	\includegraphics[width=\columnwidth]{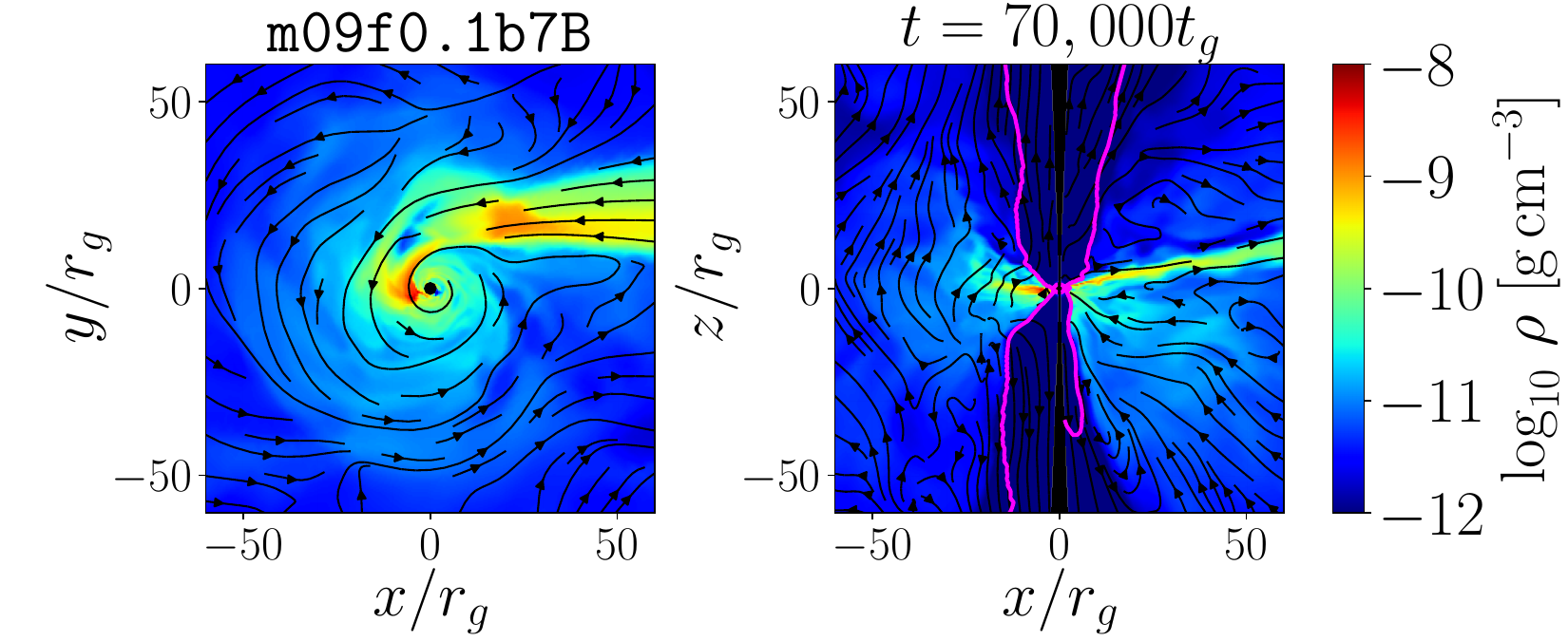}\\
    \includegraphics[width=\columnwidth]{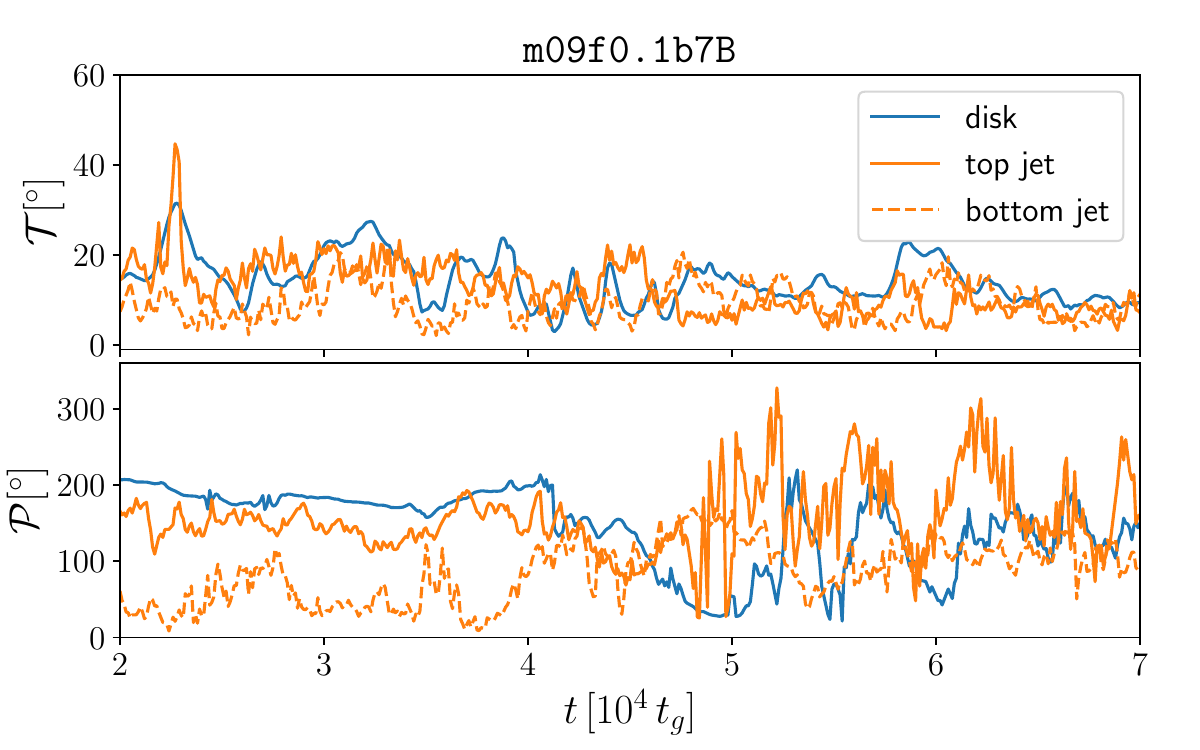}
    \caption{The same as \autoref{fig:m09_str10Edd_b7}, but for model \texttt{m09f0.1b7B} which is a restart of \texttt{m09f0.01b7} at $t = 2 \times 10^4 t_g$. $f_\rho$ is instantaneously increased from 0.01 to 0.1 at the start of the simulation. Since the change in density contrast is milder than \texttt{m09f1b7B}, the stream manages to penetrate the disk, but loses a substantial amount of orbital energy similar to model \texttt{m09f0.1b7A}. As a result, the self-intersection outflow persists, but is much weaker. The tilt of the jet and disk slowly decreases over the course of the simulation until it was observed to reach a rough equilibrium of about $10^\circ$. Although magneto-spin alignment is able to realign much of the inner system, filaments of tilted material linger in the disk which may contribute to the residual tilt in the system as well as the wild precession observed in the jet at late times.}
    \label{fig:m09_str100Edd_b7_C}
\end{figure}

\begin{figure}
    \centering{}
	\includegraphics[width=\columnwidth]{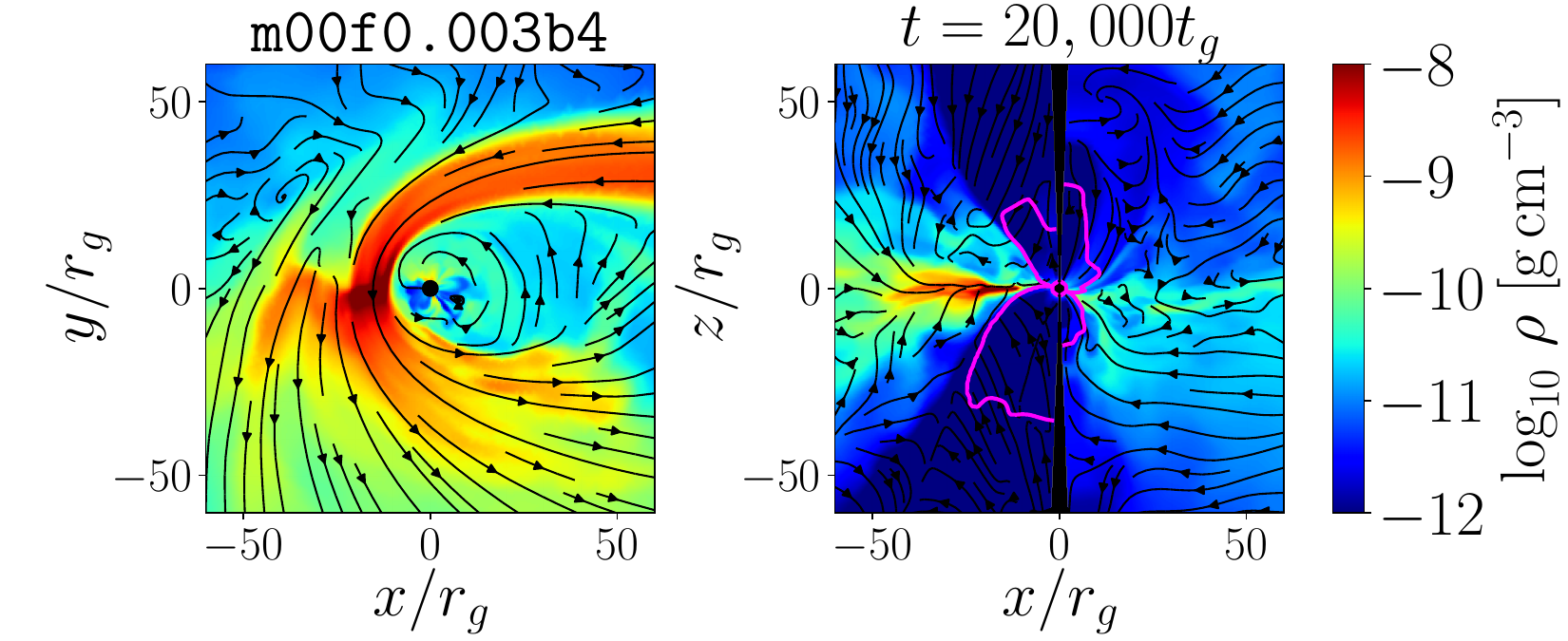}\\
    \includegraphics[width=\columnwidth]{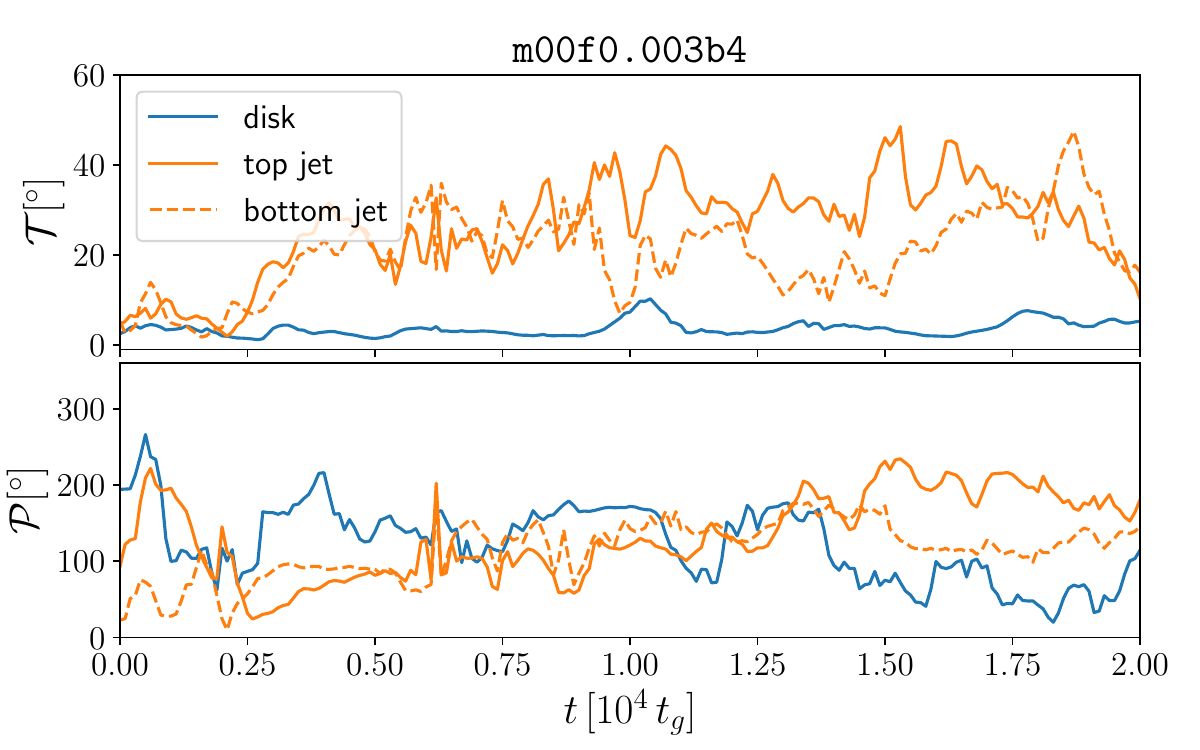}
    \caption{The same as \autoref{fig:m09_str10Edd_b7}, but for model \texttt{m09i100b4}. Note we show the initial state in the top row and the final state of the simulation in the bottom row. Since there is no jet, the corona is observed to tilt by $\mathcal{T} > 20^\circ$ due to the self-intersection outflow. However, the disk tilt remains approximately aligned with the BH spin. This confirms that a jet is necessary to induce a tilt instability in MAD TDE disks. }
    \label{fig:m00_str100Edd_b4}
\end{figure}

\begin{figure}
    \centering{}
	\includegraphics[width=\columnwidth]{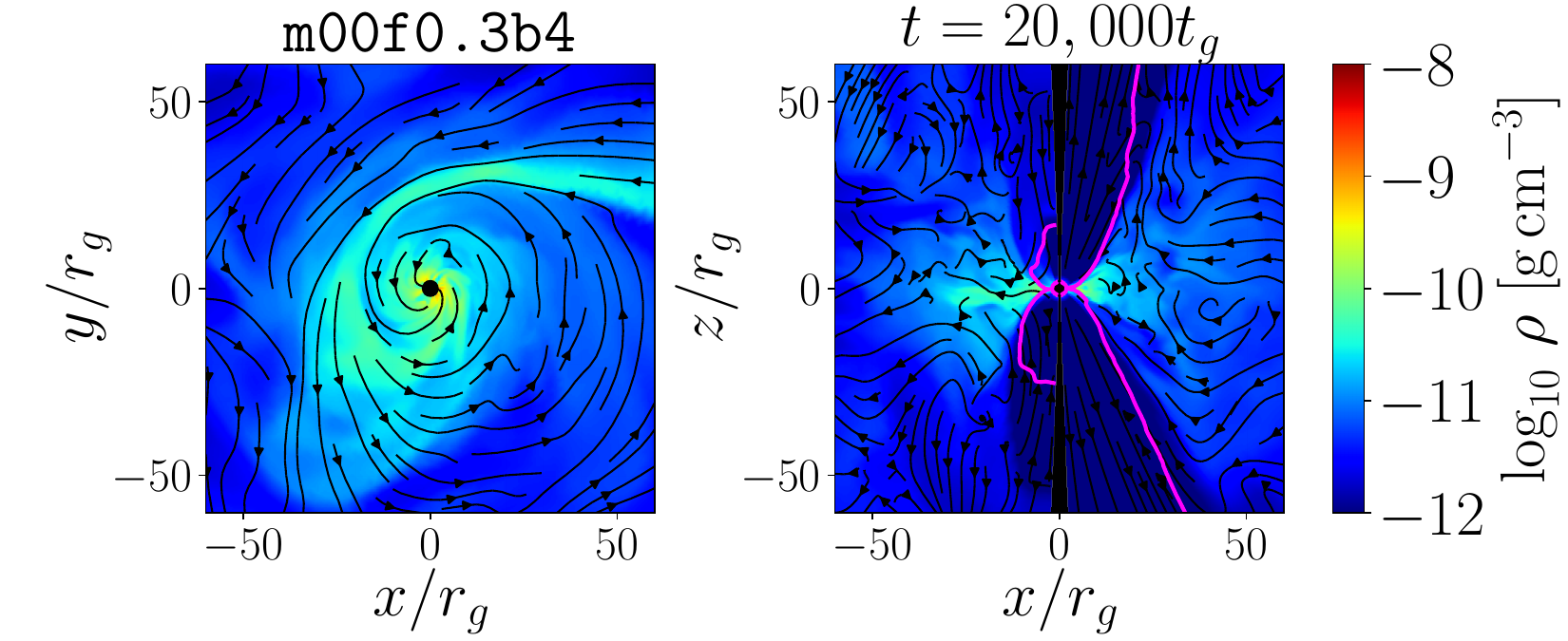}\\
    \includegraphics[width=\columnwidth]{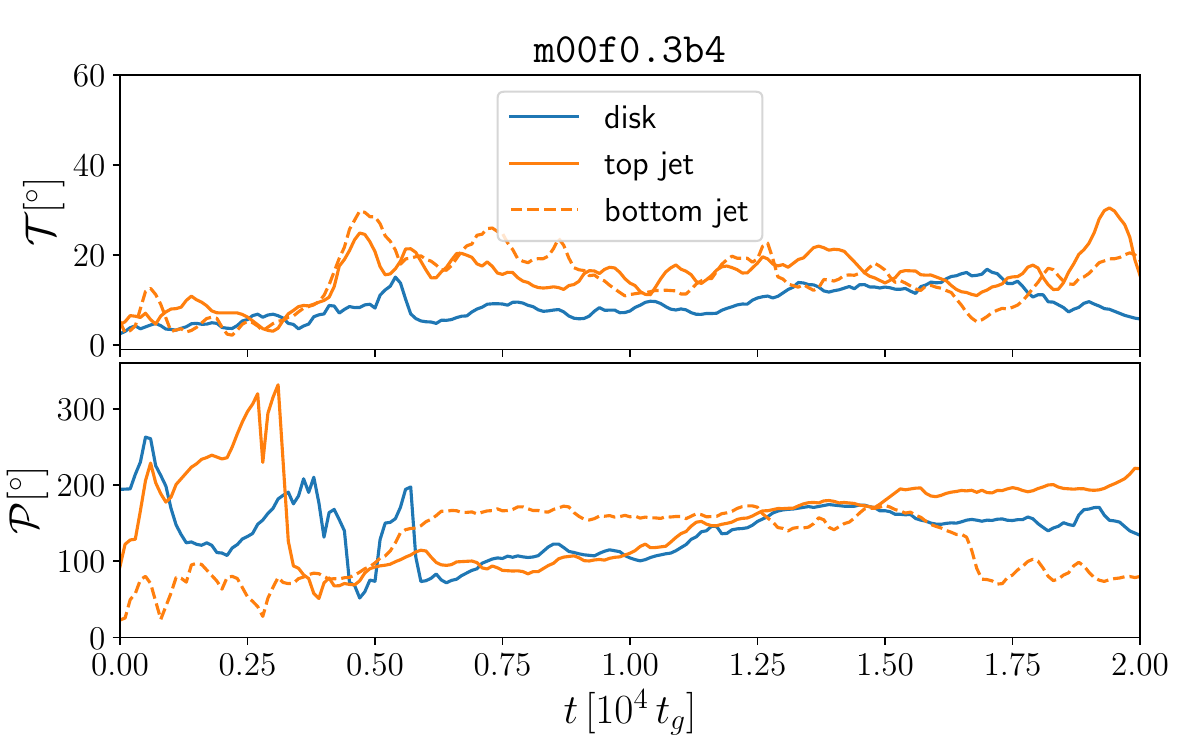}
    \caption{The same as \autoref{fig:m09_str10Edd_b7}, but for model \texttt{m09i1b4}. Due to the higher density contrast, the stream loses orbital energy on its way to pericenter, and the self-intersection outflow is negligible. Surprisingly, we measure a nonzero tilt for the corona and disk. We believe this is due to asymmetry introduced to the system by the stream in the absence of magneto-spin alignment and a strong jet.}
    \label{fig:m00_str1Edd_b4}
\end{figure}

\begin{figure}
    \centering{}
	\includegraphics[width=\columnwidth]{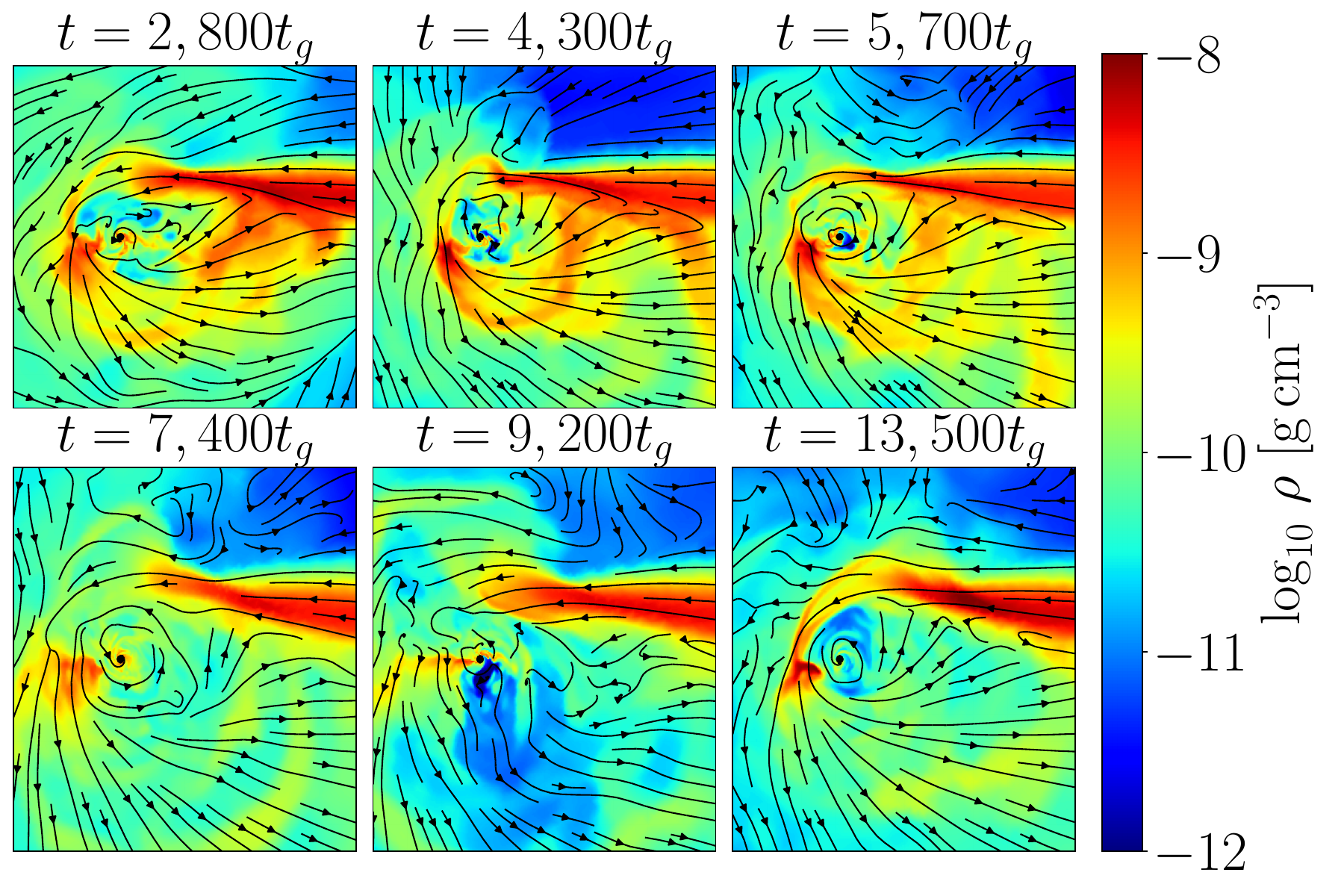}\\
    \includegraphics[width=\columnwidth]{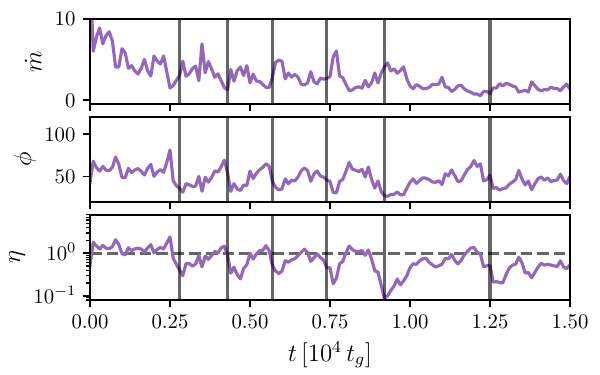}
    \caption{ Here we show snapshots of violent self-intersection events in model \texttt{m09f0.01b7} (first and second row). Colors indicate gas density and stream lines indicate gas velocity. We also show the mass accretion rate (third row), magnetic flux threading the BH (fourth row), and jet efficiency (fifth row). Vertical gray lines correspond to the same times as the snapshots shown in the first and second rows, $2,800,\,  4,300,\, 5,700,\, 7,400,\, 9,200,\, \rm{and\,} 13,500 t_g$, respectively. The violent self-intersections are accompanied by a drop in magnetic flux and jet power. We also note a small increase in mass accretion rate, which is less dramatic than the change in magnetic flux and jet efficiency.}
    \label{fig:disruptions}
\end{figure}

\begin{figure}
    \centering{}
	\includegraphics[width=\columnwidth]{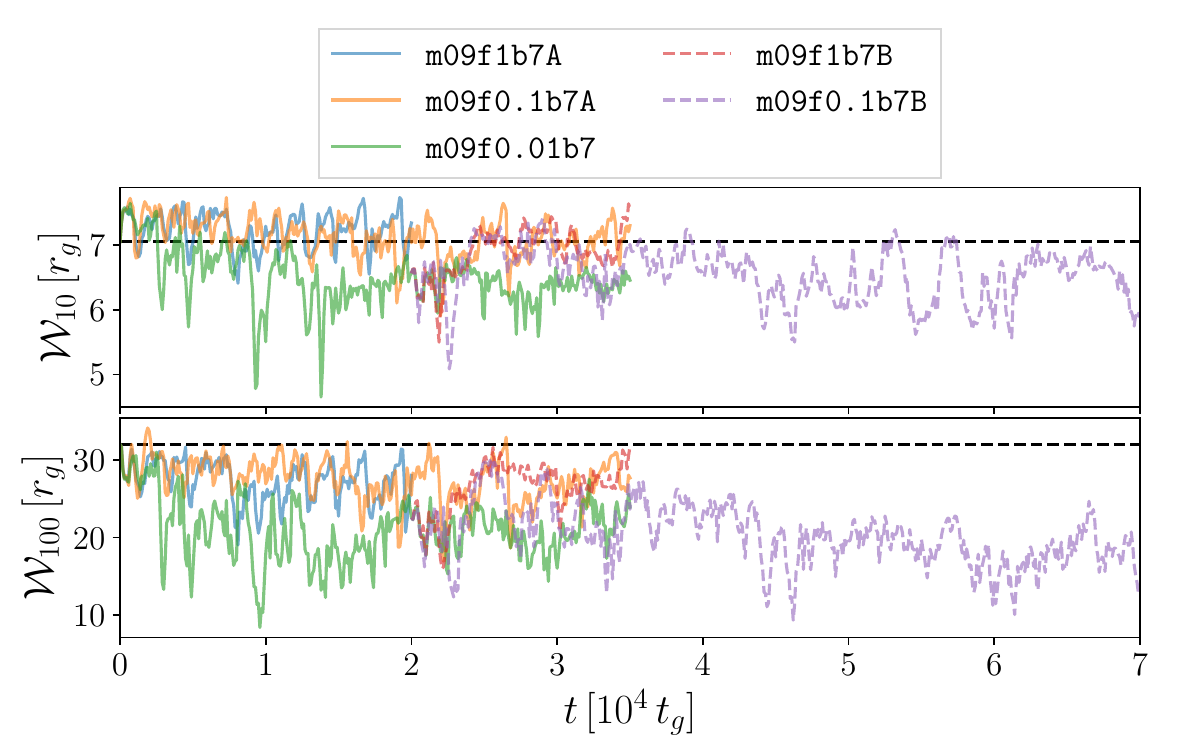}\\
 	\includegraphics[width=\columnwidth]{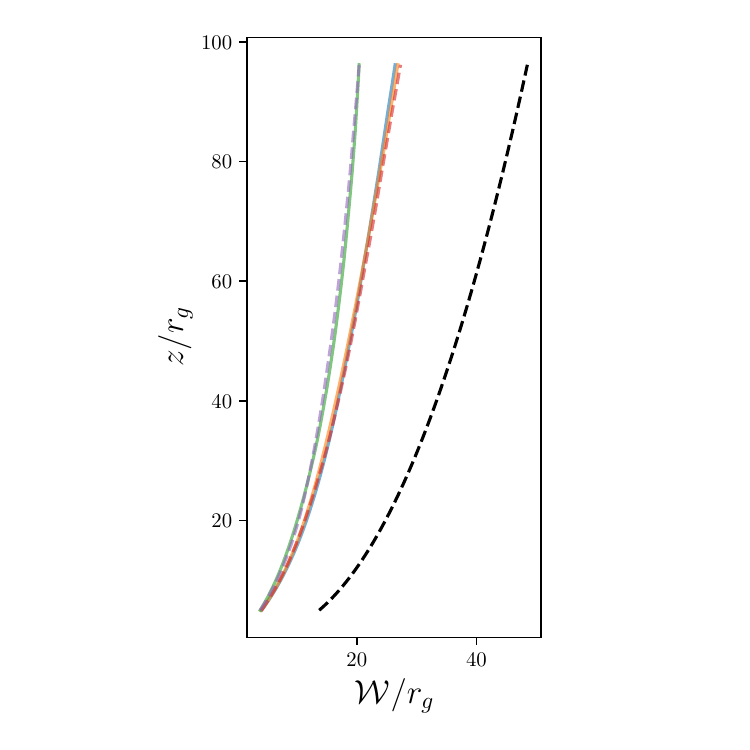}
    \caption{We show the mean jet width at $r = 10 r_g$ ($\mathcal{W}_{10}$, top panel) and $r = 100 r_g$ ($\mathcal{W}_{100}$, middle panel) as a function of time for each model. In the bottom panel we show the mean jet width as a function of $z$ and time averaged over $t_{\rm start} + 5000 t_g$ to $t_{\rm end}$. We also show the jet profile for the $a_* = 0.9$ model from \citet{Narayan2022} (dashed black line). We describe the figures in \autoref{sec:jetcollimation}. }
    \label{fig:solidangles}
\end{figure}


At the onset of stream injection, since the stream is substantially denser than the pre-existing MAD disk with $f_\rho = 0.01$, the stream is largely unperturbed by the disk material on its path towards pericenter. Subsequently, the stream precesses and violently self-intersects with itself at the self-intersection radius. 

Between $t = 0 - 0.7 \times 10^4 t_g$, the self-intersection outflow begins to tilt the jet and we measure tilt angles for both the top and bottom jet of $\sim 10^\circ - 20^\circ $. During this initial stage, the disk remains aligned with the BH spin. Between $t =  0.7 - 1.2 \times 10^4 t_g$, the disk tilt begins to increase until it roughly equals the tilt angle of the top and bottom jets. During this stage, the precession angle oscillates wildly, in part due to the initial tilt angle of zero.

For $t > 1.2 \times 10^4 t_g$, the tilt of the top jet and disk continue to grow until $\mathcal{T}_{\rm jet,top} \sim 30^\circ$ and $\mathcal{T}_{\rm disk} \sim 23^\circ$. In a typical tilted MAD disk system, the jet acts to align the inner accretion disk with the BH spin. However, once the disk tilt begins to grow in \texttt{m09f0.01b7}, it is unable to realign with the BH spin due to already tilted disk material adding angular momentum at the self-intersection radius. This sets up a tilted system which is shown to be stable for at least the duration of the simulation (roughly $\sim 2.3\times10^4 t_g$). Interestingly, the jet precession angle does not show strong variability after the disk tilts. Instead the top and bottom jets show nearly constant precession angles that are roughly $180^\circ$ out of phase at $t>2.3\times10^4 t_g$.

Volume renderings of the evolution are shown in \autoref{fig:volumerendering}. Equatorial and poloidal slices as well as the full time evolution of the tilt and precession angles are shown in \autoref{fig:m09_str100Edd_b7_A}.

\subsubsection{Medium Density Contrast Jetted Model: $f_\rho=0.1$} \label{sec:mediumfrho}

Since this model has an intermediate density contrast, the stream is still able to flow towards the BH. However, it is significantly perturbed and the pericenter radius is shifted slightly outward, which also increases the self-intersection radius. This leads to a substantially weakened self-intersection and self-intersection outflow. As a result, the jet is only slightly perturbed by the outflow and we find that the jet remains stable with $\mathcal{T} \lesssim 10^\circ$ and the disk remains aligned with the BH spin throughout the entire evolution. The precession angle is not meaningful here due to the near perfect alignment. See \autoref{fig:m09_str10Edd_b7} for visualations and the time evolution. 

\subsubsection{High Density Contrast Jetted Model: $f_\rho=1$} \label{sec:highfrho}

In this model, the density contrast is large enough that the stream experiences extreme ram pressure from the accretion disk and is halted at $r \sim 50 - 100 r_g$. The stream material never reaches pericenter and instead mixes in with the pre-existing disk. Consequently, the system resembles a standard MAD ADAF and neither the jet or disk show large changes in their tilt. Again, the precession angle is not meaningful here due to the near perfect alignment. This evolution is depicted in \autoref{fig:m09_str1Edd_b7}.

\subsubsection{Restarts of \texttt{m09f0.01b7} with Higher Density Constrast} 

For model \texttt{m09f1b7B}, we perform a restart of \texttt{m09f0.01b7} at $t = 2 \times 10^4 t_g$ with $f_\rho$ instantaneously increased from 0.01 to 1. The self-intersection is rapidly halted due to the increased density contrast and the jet subsequently realigns with the BH spin. The tilt of the disk remains slightly elevated above the tilt of the jet. This is due to the density weighting applied in \autoref{eq:Jdisk}, which gives larger weighting to higher density remnants of the tilted gas which is still in the simulation domain. However, as can be seen in \autoref{fig:m09_str100Edd_b7_B}, the inner disk is able to realign with the BH spin by the end of the simulation. We expect in a physical scenario the system will have time to adjust and the disk tilt should completely realign with the BH spin similar to the jet.  

For model \texttt{m09f0.1b7B}, we also perform a restart of model \texttt{m09f0.01b7} at $t = 2 \times 10^4 t_g$, but with $f_\rho$ instantaneously increased from 0.01 to 0.1. Similar to \texttt{m09f0.1b7A}, the stream is only perturbed from its orbit and the self-intersection still persists, but is weakened as a result. With weaker ram pressure acting on the jet, the jet and disk begin to realign with the BH spin. However, this process is much slower than in model \texttt{m09f1b7B}, and we find that the disk and jet tilt are highly variable until finally decreasing until they settle at $\mathcal{T} \sim 10^\circ$ by the end of the simulation (see \autoref{fig:m09_str100Edd_b7_C}). The total run time of the simulation (see \autoref{tab:table1}) corresponds to only roughly three days for a $10^6 M_\odot$ BH which suggests, assuming rapid transitions in the density contrast, that the tilt can evolve rapidly enough to explain features such as jet shut-off as we discuss later in this work. 

\subsubsection{Non-Jetted Models}

For the low density contrast model (\texttt{m00f0.003b4}), the initial evolution of the streams is similar to that of model \texttt{m09f0.01b7}. The self-intersection and self-intersection outflow result in a ram pressure which tilts the jet region. However, as there is no true jet since $a* = 0$, the jet region that we measure may be thought of as a corona. As shown in \autoref{fig:m00_str100Edd_b4}, the corona becomes substantially tilted with $\mathcal{T} \sim 20^\circ - 40^\circ$. The disk remains perfectly aligned with the BH spin throughout the entire evolution. This demonstrates that a powerful jet is responsible for the tilt instability that we observe in \texttt{m09f0.01b7}. 

For the low density contrast model (\texttt{m00f0.3b4}), the stream is perturbed due to its interaction with the pre-existing disk, similar to \texttt{m09f1b7A}. However, we find that the disk tilt increases slightly over the course of the simulation ($\mathcal{T} \lesssim 10^\circ$, see \autoref{fig:m00_str1Edd_b4}). The corona attains a tilt of $\mathcal{T} \sim 20^\circ$. This is due to asymmetry introduced to the system as the stream interacts with the disk. Since the magnetic field is strong, and stream material cannot steadily feed aligned material to the inner disk, the tilted corona is capable of tilting the disk. Unlike model \texttt{m09f1b7A}, magneto-spin alignment does not counteract any induced tilt. Tilt induction in a MAD around a non-spinning BH was also demonstrated by \citet{Ressler2020} in the context of a stellar wind fed model of Sagittarius A$^*$, suggesting tilt induction may be common in MAD disks around non-spinning BHs that are fueled by asymmetrical inflows.

\subsection{Violent Self-Intersections and Variability} \label{sec:variability}

For the first $15,000\, t_g$ of model \texttt{m09f0.01b7}, we identify six complete stream disruptions at times $(2,800,\,  4,300,\, 5,700,\, 7,400,\, 9,200,\, 13,500)t_g$ as shown in \autoref{fig:disruptions}. These correspond to a temporal separation of $(1500,\, 1400,\, 1700,\, 1800,\, 3300)t_g$. Assuming a Keplerian orbit, this corresponds to an orbital radius of $(38,\, 37,\, 42,\, 43,\, 65)r_g$. These are similar to the self-interaction radius of $\sim 50 r_g$, which is to be expected in the case of a feedback loop caused by angular momentum transfer during self-intersection \citep{Sadowski2016, Curd2021}.

Here we find that not only does the mass accretion rate vary during these events, but the magnetic flux threading the BH drops from $\phi_{\rm{BH}}\sim60$ to $\sim40$. Since the disk is MAD and the BH is rapidly rotating, this will inevitably lead to flaring behaviour. Indeed, we see the total efficiency drop from $\sim100\%$ at the peaks to $10-50\%$ at the minima. We discuss how our model can be applied to the variability in jetted TDEs like \textit{Swift} J1644+57 in \autoref{sec:discussion}.

\subsection{Jet Collimation} \label{sec:jetcollimation}

We measure the mean jet width at $r = 10 r_g$ ($\mathcal{W}_{10}$) and $r = 100 r_g$ ($\mathcal{W}_{100}$) as a function of time following \autoref{eq:wjet} in \autoref{fig:solidangles}. The jet width shows oscillations as a function of time due to the highly variable magnetic flux. This is typical of a MAD disk, but here we are focused on the average behavior of the jet. 

For model \texttt{m09f0.01b7}, the self-intersection outflow causes substantial collimation. The velocity stream lines in the right middle panel of \autoref{fig:m09_str100Edd_b7_A} show high density material sometimes flowing completely in the $-x$ direction which will provide substantial ram pressure on the jet. We see a decrease of roughly $10 r_g$ in the jet width measured at $100 r_g$ compared to the initial jet. 

For models \texttt{m09f1b7A} and \texttt{m09f0.1b7A}, the jet width at $r = 10 r_g$ is similar to that of the initial jet prior to injection due to the weakening of the self-intersection outflow. However, we do observe slightly more collimation at $r = 100 r_g$ compared to the initial jet, perhaps due to changes in the outflow properties when the stream interacts with the disk. For instance, the velocity stream lines in \autoref{fig:m09_str1Edd_b7} and \autoref{fig:m09_str10Edd_b7} show flows towards the jet axis, which are not present in the initial jet (see top panel of \autoref{fig:m09_str100Edd_b7_A}). This may lead to more collimation in TDE jets as they propagate outwards compared to a standard MAD; however, we limit ourselves to measuring the jet profile for $r \leq 100 r_g$ due to poor angular resolution of the jet at larger radii. 

For model \texttt{m09f1b7B}, once the self-intersection ceases due to the increased $f_\rho$, the jet width returns to near the initial value within $\sim 5000 t_g$. However, model \texttt{m09f0.1b7B} shows a much narrower jet when compared with model \texttt{m09f0.1b7A}. This is not due to the self-intersection outflow, but the magnetic flux dropping off towards the end of the simulation. 

We also time average the jet width from $t_{\rm start} + 5000 t_g$ to $t_{\rm end}$ (see bottom panel in \autoref{fig:solidangles}). We find similar jet profiles for all models with weak self-intersection outflows (\texttt{m09f1b7A}, \texttt{m09f0.1b7A}, \texttt{m09f1b7B}). Model \texttt{m09f0.1b7B} is similar to model \texttt{m09f0.01b7}, but again this is due to a decrease in magnetic flux and not a result of the self-intersection outflow. We compare our results with the jet profile for the $a_* = 0.9$ model from \citet{Narayan2022}. We find that our initial conditions result in a slightly narrower jet, but the profile appears to be quite similar for the models with weak self-intersection. 

\subsection{Gas Temperature} \label{sec:temperature}

\begin{figure*}
    \centering{}
	\includegraphics[width=0.5\textwidth]{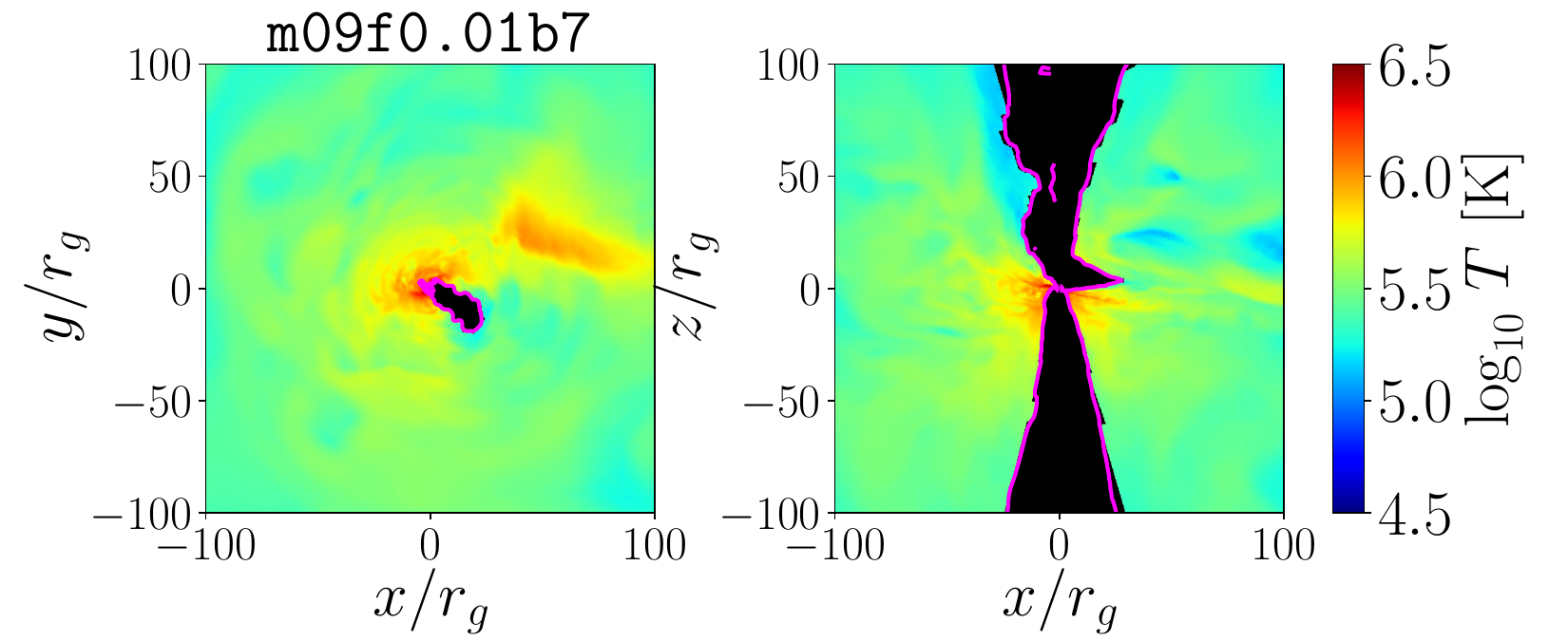}\\
 	\includegraphics[width=0.5\textwidth]{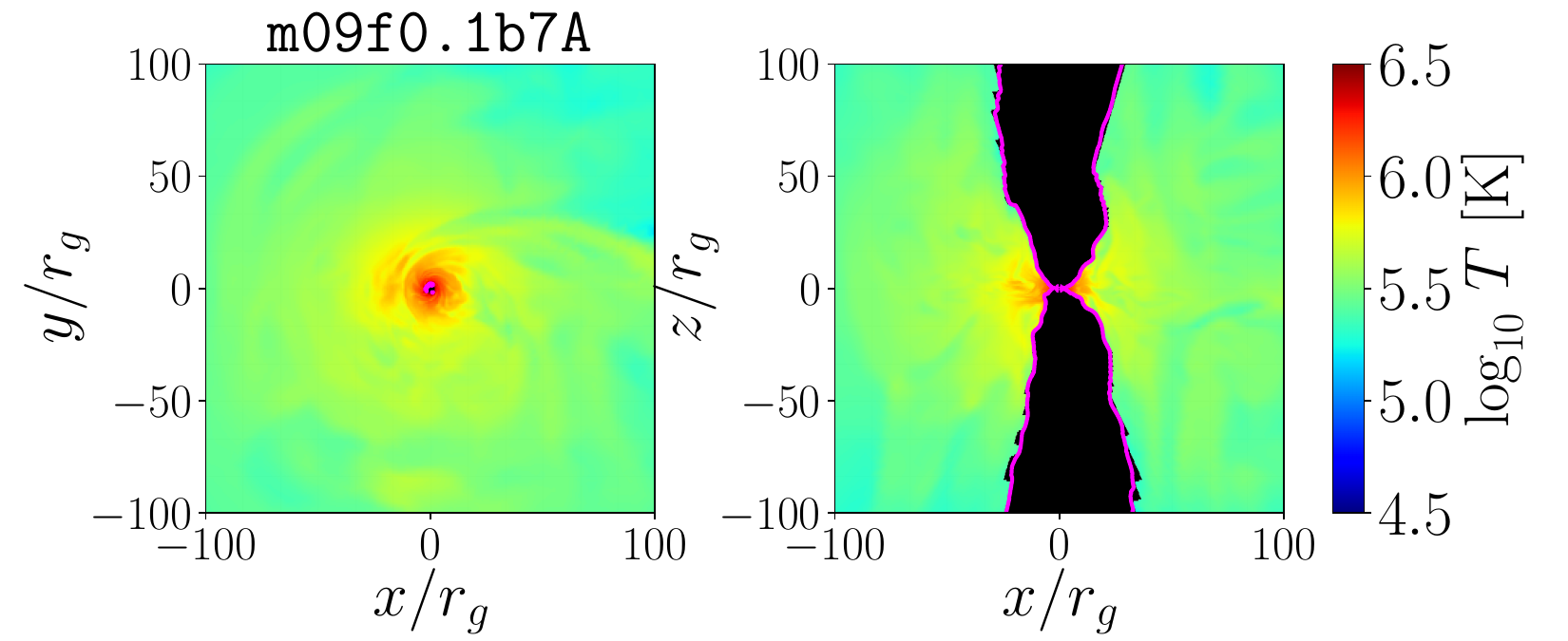}\\
 	\includegraphics[width=0.5\textwidth]{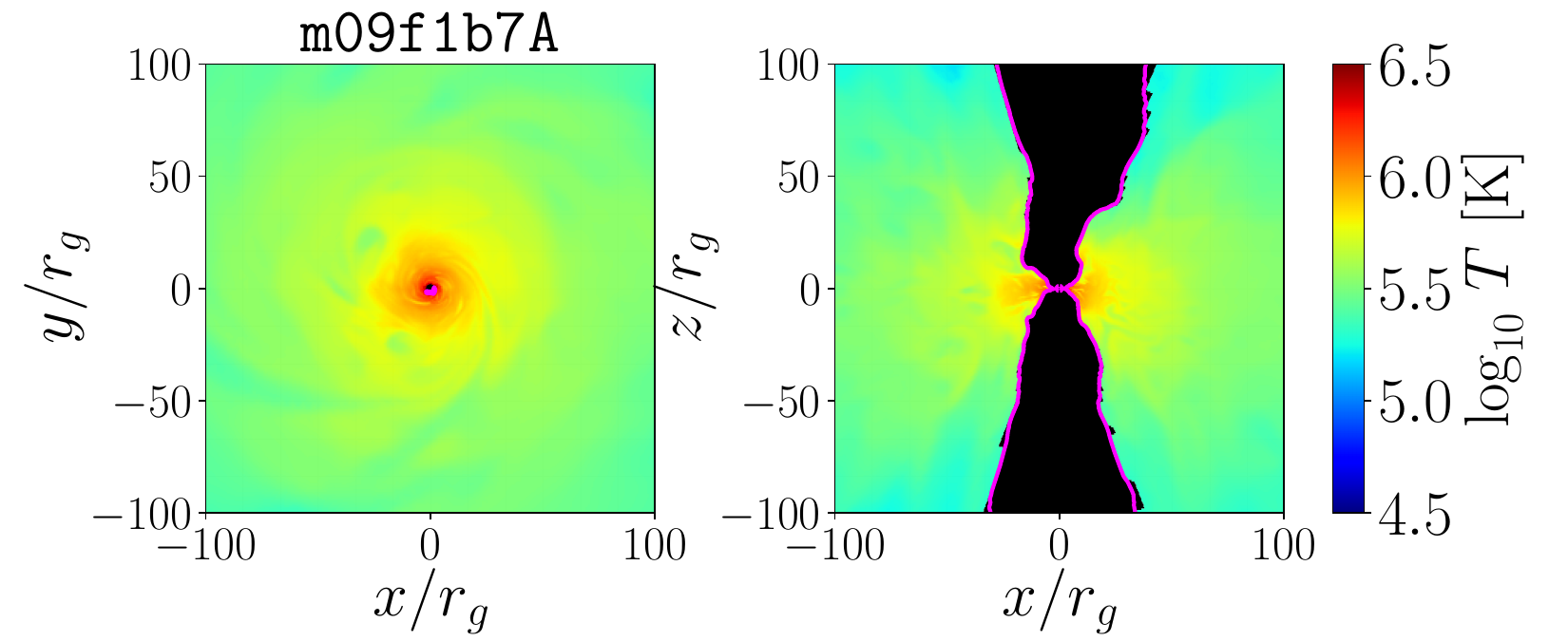}\\
   	\includegraphics[width=0.5\textwidth]{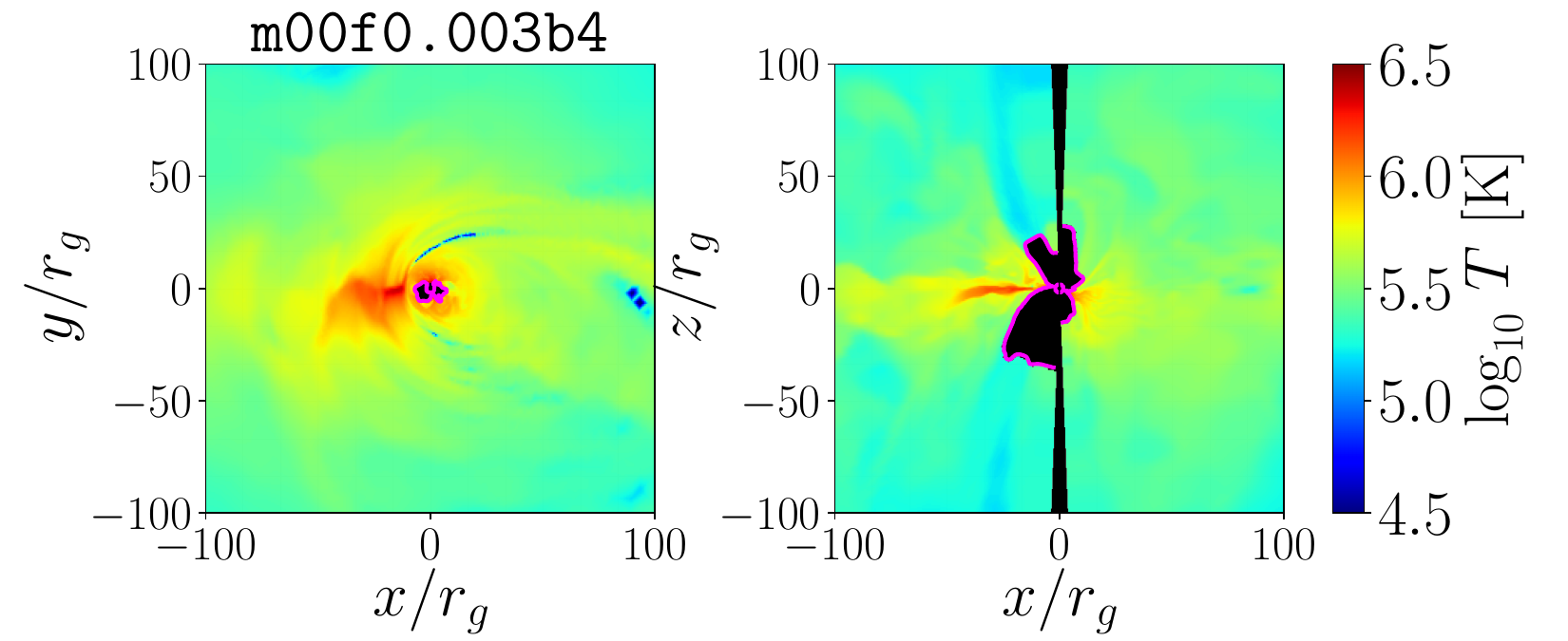}\\
   	\includegraphics[width=0.5\textwidth]{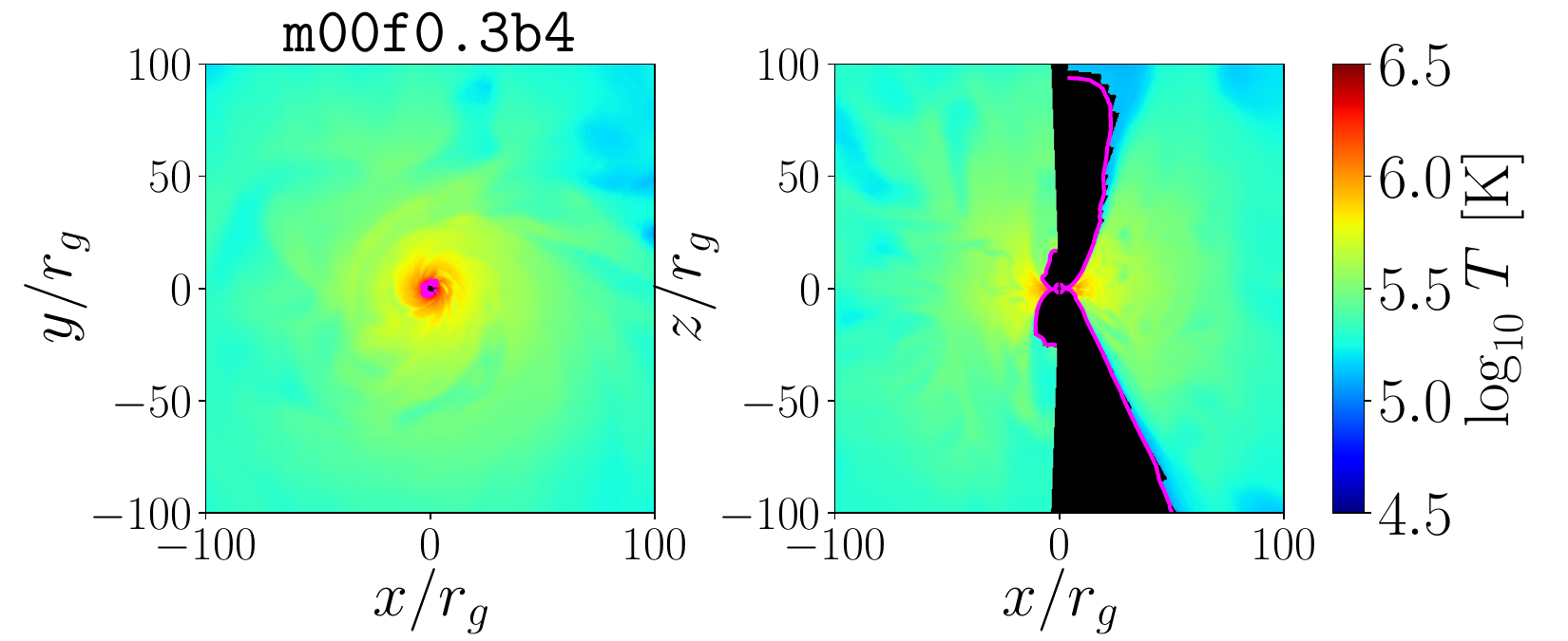}
    \caption{Here we show the gas temperature (colors) and $\sigma=1$ boundary (pink line) for each model at the final snapshot. We mask the gas temperature in regions where $\sigma > 1$ for numerical reasons. } 
    \label{fig:gastemperature}
\end{figure*}

We estimate the gas temperature in the disk by accounting for radiation under the assumption that the disk is optically thick. We split the temperature into gas and radiation by solving 
\begin{equation}
    p_g = \dfrac{\rho k T}{\overline{m}} + \dfrac{1}{3}a T^4 ,
\end{equation}
where $\overline{m}$ is the mass per particle and $T$ is the temperature. The gas temperature in the $\sigma>1$ region is uncertain due to both numerical floors and the use of entropy inversion when energy conserving inversion fails in highly magnetized zones in GRMHD. As a result, we mask the gas temperature in the jet/corona, but we generally expect it to be substantially hotter than the disk \citep{Curd2019}. We show the gas temperature for each model at $t=t_{\rm end}$ in \autoref{fig:gastemperature}.

In the accretion disk, since the gas and radiation pressure are split evenly, the gas temperature of the accretion disk reaches $T \sim 10^{5-6}$ K, which approximately agrees with \citet{Curd2019}. Nozzle and self-intersection shocks also contribute to heating the gas and drive the temperature up to $\sim 10^{6}$ K at radii up to $50-100r_g$.

In models with a prominent jet, the gas temperature may exceed $10^{6}$ K where $\sigma>1$, which is in the range for X-ray photon production \citep{Curd2019}. Since the jet is able to prevent polar inflows, the poles will remain optically thin even at the peak of the fallback rate, allowing jet emission to emerge. Comptonization within this region is expected to produce a hard spectrum which shines even in the $\gamma$-ray band. 

The non-jetted models on the other hand may have their X-ray emission largely absorbed if the photosphere is roughly spherical early on. Only after the funnel can form (or the photosphere recedes) can X-rays emerge.

\section{Discussion} \label{sec:discussion}

\subsection{Variability Driven by Violent Self-Intersection}

\textit{Swift}J1644+57 showed variability on a range of timescales with both short period QPOs at $\sim 200 \rm{s}$ \citep{Reis2012} and long period dips in the light curve on time scales of $\sim 10^6 \rm{s}$ \citep{Saxton2012}. The short period QPOs are thought to originate from short term variability on the horizon scale due to orbits or resonances in the inner accretion disk. The long period variability has been suggested to arise from wobbling of the jet \citep{Tchekhovskoy2014}, periodic violent stream self-intersection \citep{Andalman2022}, or magnetic flux eruption events \citep{Curd2023}.  

Previous global simulations of forming TDE disks have identified complete disruptions of the incoming stream in cases where $\beta=3-7$ \citep{Curd2021, Andalman2022}. The disruptions are temporally spaced by roughly the orbital period at the self-intersection radius. The fact that such a periodic dynamical effect took place was viewed as an attractive explanation for the variability in J1644+57. However, with no magnetic fields or radiative transfer calculations available, \cite{Andalman2022} hypothesized that this interaction could drive flaring through changes in the mass accretion rate at the horizon.

As shown in \autoref{sec:variability}, we directly relate the  complete disruptions during self-intersection with jet variability. Since the total efficiency in \autoref{fig:disruptions} correlates directly to jet power in MAD accretion disks, this can account for the large drops in the flux seen in J1644+57, which had minima as low as $\lesssim50\%$ of the maximum. This is solid confirmation of the idea proposed by \cite{Andalman2022}; however, we suggest that flaring is not because of changes in the mass accretion rate directly. Rather, it is the fact that the stream acts to keep magnetic flux anchored to the BH. The magnetic flux threading the BH is at the saturation value prior to when the stream disrupts itself during self-intersection. When the feeding from incoming steam material is temporarily halted, magnetic flux eruptions shed flux until $\phi$ settles to a lower value. The disk injection simulations presented in \cite{Curd2023} found that after flux eruption events the magnetic flux took roughly the orbital period at the injection radius to recover. This is dynamically similar to the effects seen in this work; however, here the period is directly related to the self-intersection radius rather than the gas injection radius.

Given the relationship between the variability period and the self-intersection radius, this suggests that X-ray variability can be related to the orbital properties of the disrupted star in a jetted TDE. For instance, assuming $M_{\rm BH} = 10^6 M_\odot$ for J1644+57, the roughly $10^6$ second variability corresponds to a self-intersection radius on the order of $10^3 r_g$. For an $a_* = 0 $ BH, this corresponds to a $\beta \sim 1.5$ TDE. The steady increase in the variability period may be due to an increase in the self-intersection radius as the disk builds up over time as illustrated by \citet{Ryu2023}.

We will explore the properties of magnetized TDE disks and magnetic flux saturation in more detail in a future report.

\subsection{Tilt Evolution in Jetted TDEs}

Our simulations illustrate that even an aligned TDE can undergo strong tilt excitation when a jet is present. The fact that the tilt decreases when the density contrast increases, which is due to the self-intersection shock and outflow being weakened, suggests that X-ray shut-off in TDE jets may be possible even without the disk exiting the MAD state. A precise determination of the dependence of $\mathcal{T}_{\rm jet}$ on $f_\rho$ would require many more simulations or a semi-analytic prescription of the jet dynamics. However, we can infer that there is a critical value of $f_\rho$ above which the tilt remains aligned with the BH spin axis.

The X-ray variability in \textit{Swift} J1644+57 indicates that X-rays originate from near the BH, so changes in the jets orientation at $r<100 r_g$ are expected to directly impact the beamed X-ray flux. In the off-axis jet model developed by \citet{Beniamini2023}, the off-axis jet flux is proportional to the on-axis jet flux through a simple beaming correction factor
\begin{equation} \label{eq:beaming}
    a=\dfrac{1 - \beta_{\rm{jet}}}{1 - \beta_{\rm{jet}}\cos(\mathcal{T}_{\rm{obs}}-\mathcal{T}_{\rm{jet}})},
\end{equation}
where $\beta_{\rm{jet}}$ is the jet velocity $\mathcal{T}_{\rm{obs}}$ is the angle of the observer relative to the $z$-axis. The flux is approximated as
\begin{equation} \label{eq:Fmodel}
    F(\mathcal{T}_{\rm{jet}})=F_{\rm{on,jet}}(t)
     \begin{cases}  
       1 , \quad  & \Delta\theta  < \theta_{\rm{jet}} \\
       0.5 a^2 , \quad  &  \theta_{\rm{jet}} < \Delta\theta  < 2\theta_{\rm{jet}} \\
       0.5 a^3 , \quad  & \Delta\theta  > 2\theta_{\rm{jet}}.
     \end{cases}
\end{equation}
Here $\Delta\theta\equiv\mathcal{T}_{\rm{obs}}-\mathcal{T}_{\rm{jet}}$ and $\theta_{\rm{jet}}=\gamma_{\rm{jet}}^{-1}$ is the angle that the jet is beamed into. The factor of $0.5$ is a geometrical correction.

For a \textit{Swift} J1644+57 like jet, we assume initial values of $\gamma_{\rm{jet}}=10$, $\beta_{\rm{jet}}\approx 0.995$, $\theta_{\rm jet}\approx 6^\circ$, and $\mathcal{T}_{\rm{jet}}=20^\circ$. Assuming the observer is aligned with the jet initially with $\mathcal{T}_{\rm{obs}}=20^\circ$, it can be seen from \autoref{eq:Fmodel} that a shift of $\mathcal{T}_{\rm{jet}}=20^\circ$ to $\mathcal{T}_{\rm{jet}}=0^\circ$ would correspond to a roughly 3 orders of magnitude decrease in flux (since $a\approx0.077$ for our assumed jet parameters). Even a more modest tilt change of $\mathcal{T}_{\rm{jet}}=20^\circ$ to $\mathcal{T}_{\rm{jet}}=10^\circ$ would give a nearly 2 orders of magnitude decrease (since $a\approx0.24$). In short, jet realignment and relativistic beaming effects could account for the roughly 2 orders of magnitude drop in X-rays at $\sim500$ days in less than 15 days in \textit{Swift} J1644+57. This also applies to other jetted TDEs.

Note that we only require that the X-ray emission decrease by at least two orders of magnitude within $\sim2$ weeks in order to explain the behaviour of \textit{Swift} J1644+57. The X-rays after the decline could be disk emission which becomes dominant when the jet is out of the line of sight. This is more attractive of an explanation since the jet emission follows a $t^{-5/3}$ power law even after tilting, but the late time X-rays are approximately constant.

While jet tilt can explain changes in X-ray flux in jetted TDEs, rapid changes in the jet tilt still require a rapid mechanism of some kind. Without more complex modeling, we cannot make precise statements on different mechanisms. However, we suggest two candidates.

The first is disk state transitions. Accretion disks can have both thick and thin regions as a function of radius due to different contributions of radiative and advective cooling \citep{Sadowski2011}. Furthermore, the innermost disk could remain thermally stable due to magnetic pressure support even as the accretion rate becomes sub-Eddington if the disk is MAD \citep{Curd2023}. Meanwhile, the outer disk could become thermally unstable as the mass fallback rate becomes near-Eddington. Vertical collapse of the outer disk would rapidly increase $f_\rho$ by more than an order of magnitude due to the disk density having a $h_d^{-1}$ scaling. In turn, the self-intersection would weaken, and the jet tilt would align with the BH spin axis. Although our simulations use geometrically thick MADs, the rapid change in $f_\rho$ for models \texttt{m09f0.1b7B} and \texttt{m09f1b7B} is synonymous with a vertical disk collapse, since $f_\rho$ drives the stream-disk interaction.

The second is the self-intersection radius itself. We only had sparse coverage of $f_\rho$ in this work due to the cost of each simulation. It is possible that the self-intersection outflow weakens rapidly enough as $f_\rho$ increases to explain rapid X-ray flux evolution without relying on disk collapse. In bent radio jets, the ballistic model relates the jet angle to a force balance between centrifugal forces and the ram pressure from the external medium. In TDEs, the ram pressure of the self-intersection outflow roughly scales with the self-intersection radius as $r_{\rm SI}^{-5/2}$. Here we have assumed the self-intersection outflow has a constant velocity upon launching, the velocity is the escape velocity, and the outflow has a spherical geometry. In principle, such a strong dependence of the ram pressure on the self-intersection radius could lead to rapid changes as $f_\rho$ increases since this pushes $r_{\rm SI}$ further out (i.e. see \autoref{fig:Large_Scale_maps}); however, this needs to be studied in detail.

\subsection{Coronal Evolution in Non-Jetted TDEs}

Tilt effects are unlikely to lead to substantial X-ray changes in non-jetted TDEs since the emitting region is non-relativistic and we only saw changes of up to $\sim10^\circ$ in our models. However, our $a_*=0$ simulations demonstrate that a coronal region can be sustained even during stream self-intersection provided enough magnetic flux threads the disk/BH. \citet{Curd2021} found no funnel/corona region during the stream injection phase due to a substantially lower magnetic flux than our MAD disks, but this may only apply to the TDE evolution near the peak fallback rate as their simulations covered only a few days of evolution. Assuming magnetic flux increases as a function of time, which appears to occur in TDE disks \citep{Sadowski2016}, our $a_*=0$ simulations may be interpreted as the limiting state for a TDE at a given $f_\rho$ around a Schwarzschild BH since they are MAD. Increases in X-ray emission in non-jetted TDEs may then be related to both a hot, magnetized corona forming as $\phi$ increases combined with a decrease in optical depth as the fall back rate declines. The X-rays during this phase would exhibit a slow rise as the photosphere radius drops.

The X-ray emission in AT2021ehb steadily turned on until it reached a maximum of $\sim5\times 10^{43} \rm{erg\, s^{-1}}$ before promptly declining by an order of magnitude at $\sim 270$ days \citep{Yao2022}. The rise phase and spectral hardening of AT2021ehb could be explained by the coronal evolution scenario outlined in the previous paragraph while the rapid decrease in X-ray flux could conceivably be due to disk collapse. While the coronal evolution in our non-jetted models is expected to be similar to a non-MAD case, whether or not thermal TDEs are also MAD is unclear and simulations which evolve the magnetic field suggest they should not be. This leads to important dynamical differences when considering the evolution of the disk. While a MAD disk may remain magnetic pressure supported, non-MAD accretion disks are expected to become thermally unstable once pressure support is lost.

\subsection{Future Prospects}

The discovery of tilt instability in TDE disks could have profound consequences on the emission properties beyond the X-ray emission from the jet or corona. It is conceivable that the polarization signal of the disk and jet will be impacted by changes in the tilt of the system.

Although we found some evidence of enhanced normalized magnetic flux in model \texttt{m09f0.01b7} compared to a typical MAD disk (which generally show $\phi\approx40-50$), the turbulent dynamics near the horizon may have impeded this effect. The onset of the disk tilt also seems to correspond with a decrease of the normalized magnetic flux at the horizon. Simulations with the self intersection radius further away from the horizon may allow higher normalized magnetic flux to be sustained. This may lead to a normalized magnetic flux $\phi>50$ for much of the transient, which is much higher than the MAD limit assumed by \citet{Tchekhovskoy2014}.

\citet{Curd2022,Curd2023b} investigated the morphology and radio spectra of jets from SANE super-Eddington accretion disks. Such an analysis could similarly be carried out on MAD TDE disks and would provide useful insight into how the dynamics of the system effect the ultimate jet properties. We plan to investigate this in a future work. 

\section{Conclusions} \label{sec:conclusions}

\begin{itemize}
    \item All of our simulations maintained a significant magnetic flux threading the horizon even after interacting with the injected stream. Each simulation reached a MAD or semi-MAD state. Powerful jets were launched for $a_*=0.9$ models. This suggests that TDEs can remain MAD and launch spin-powered jets, provided enough magnetic flux can be advected towards the black hole during the TDE.
    \item The strength of the self-intersection outflow depends on the density ratio between the stream and the surrounding debris, $f_\rho$.  As the stream becomes less dense, ram pressure from the debris can effectively brake the stream and it eventually joins with the debris cloud/disk with either a weak self-intersection or no self-intersection at all.
    \item We found that the Maxwell stress is subdominant to hydrodynamic sources of viscosity at all values of $f_\rho$ investigated in this work. Instead, shocks and hydrodynamic viscosity drive angular momentum transport. 
    \item During the early stages of a TDE, the stream is much denser than the debris cloud/disk with $f_\rho < 0.01$ since most of the mass has yet to fallback. The stream is essentially unperturbed by the debris cloud/disk at this stage and has a strong self-intersection shock since it maintains its orbital energy. The self-intersection outflow pushes on the jet/corona region. This tilts the jet/corona by $10-40^\circ$ in our simulations. As $f_\rho$ increases, the self-intersection shock weakens and powerful jets remain aligned with the BH spin. 
    \item Because the jet is tilted by the self-intersection outflow, the jet can transfer momentum to the disk, which tilts the disk to $\sim20-30^\circ$ in less than $10,000 t_g$. This configuration is stable due to the self-intersection of tilted material within $R_{\rm{SI}}$ with un-tilted material being brought in from the stream. This effect does not occur when there is no self-intersection outflow (the stream is not dense enough) or there is no jet (as shown by our $a_*=0$ models). This suggests that jetted TDEs are tilt unstable systems.
    \item When we lowered the stream density in a restart of the model \texttt{m09f0.01b7} after the disk/jet was tilted, we found that a MAD or semi-MAD state leads to alignment of the disk/jet similar to GRMHD simulations of tilted disks. We propose that this is due to the weakening/absence of the self-intersection, which acts to maintain the tilt once it sets in. 
    \item Rapid changes in the jet tilt angle, which may occur due to the outer disk collapsing or the growth of the self-intersection radius, may explain the rapid X-ray shutoff in jetted TDEs. Jet realignment with the BH spin in models \texttt{m09f1b7B} and \texttt{m09f0.1b7B} represents a change of $\sim20-30^\circ$ in the jet angle in less than three days. We propose that this is an alternative method of rapidly dropping the X-ray flux in \textit{Swift} J1644+57 in $\sim15$ days without also requiring that the system no longer be MAD.
\end{itemize}

\section{Acknowledgements}

We thank the anonymous referee for useful comments which significantly improved the quality of this manuscript. We thank Enrico Ramirez-Ruiz for useful suggestions during preparation of the manuscript. We thank Aviyel Ahiyya for assistance with obtaining observational data. This work was supported by a grant from the Simons Foundation (00001470, BC, RA, HW, and JD). Richard Jude Anantua was supported by the Oak Ridge Associated Universities Powe Award for Junior Faculty Enhancement. Computational support was provided via ACCESS resources (grant PHY230006).

\section{Data Availability}
The data underlying this article will be shared on reasonable request to the corresponding author.

\bibliographystyle{mnras}
\bibliography{main}

\appendix

\section{Magnetic Flux in Jetted TDEs}
\label{sec:appA}

The discovery of the jetted TDE \textit{Swift} J1644+57 sparked the idea that MADs could apply to jetted TDEs \citep{Tchekhovskoy2014}. However, the magnetic flux required at the peak jet power
\begin{equation}
    \Phi_{\rm{jet}}\gtrsim 8\times10^{29}{\rm{\ G \ cm^{2}}}m_6 \left(\dfrac{L_{\rm{iso}}}{10^{48} \rm{ erg \ s^{-1}}}\right)^{1/2}f_{-3}^{1/2},
\end{equation}
is immense \citep[][Equation (6)]{Kelley2014}. Here $M_{\rm{BH}}=m_6 10^6\, M_\odot$ is the BH mass, $L_{\rm iso}$ is the isotropic equivalent luminosity, and $f=f_{-3}10^{-3}$ is the beaming factor. Adopting $m_6=1$, $L_{\rm{iso}}=10^{48} {\rm{ erg \ s^{-1}}}$, and $f_{-3}=1$, the minimum flux required for a MAD state is $\Phi_{\rm{jet}}\gtrsim 8\times10^{29}{\rm{\ G \ cm^{2}}}$. Where could such a powerful flux come from?

The obvious first choice would be from the disrupted star itself, but this hypothesis has two major impediments. The first is that the typical surface magnetic field strength is only $10-100$ G for Sun-like stars with a magnetic flux of $10^{23-25} {\rm{G \ cm^2}}$. Second, the vertical magnetic field strength drops drastically due to flux freezing as the star gets tidally disrupted. This leads to a predominantly toroidal field, making any initial poloidal flux accumulation from the stream unlikely to achieve a MAD state.

\citet{Tchekhovskoy2014} proposed that magnetic flux from a MAD fossil disk surrounding the BH prior to the star being disrupted could power the jet. 
In a MAD, every $\Delta r\sim 30(h_d/0.3)r_g$, where $h_d$ is the disk scale-height, contains as much magnetic flux as the BH. As a result, $\Phi^{\rm{MAD}}_{\rm D}\propto r$. \cite{Kelley2014} demonstrated that this fossil disk flux can be transported by the stream material as it plows through the pre-existing disk. As a result, flux out at the stream apocenter $r_a$ will be transported from $r_a$ towards $r_H$ on the order of the fallback time. Additionally, since $r_a\propto (t/t_{\rm{fb}})^{2/3}$ more magnetic flux may be transported by the stream as the TDE evolves. Since MADs produce powerful jets even during the sub-Eddington state, this model requires that any prior emission meet pre-outburst limits. For a $m_6=1$ BH disrupting a main sequence $m_*=1$ star, \cite{Tchekhovskoy2014} give constraints on the fossil disk of accretion rate $\dot{m}\gtrsim 1.6\times10^{-3}$, jet power $P_{\rm{jet}}\gtrsim 2.3\times10^{41} {\rm{\ erg \ s^{-1}}}$, and beaming corrected X-ray luminosity $L_{\rm{X,obs}}\gtrsim 7\times10^{42} {\rm{\ erg \ s^{-1}}}$. This is below the \textit{ROSAT} detection limits of $L_{\rm{X}}\lesssim 10^{44}{\rm{\ erg \ s^{-1}}}$, so the fossil disk emission would not be detectable.

Another alternative source of magnetic flux in jetted TDEs is a turbulent dynamo \citep{Krolik2011,Krolik2012}. In this scenario, the dynamo process in the turbulent accretion flow resulting from the disruption would naturally lead to generation of vertical flux. Although the net vertical magnetic flux is conserved, if patches of ordered magnetic field are dragged towards the BH, this can trigger the MAD state. This scenario was only recently demonstrated to be possible by \citet{Liska2020} in an accretion disk with an initially toroidal field thanks to GPU accelerated GRMHD simulations. It is important to note that since this process is expected to be random, there is some doubt that it sources magnetic flux in jetted TDEs since the jet would turn on and off due to temporal sign changes in the magnetic field threading the BH. However, \cite{Liska2020} demonstrated a sustained MAD state over $1.3\times10^5 t_g$, or $\sim 7$ days for $m_6=1$, so it is not yet clear if the random variability expected theoretically will occur in BH accretion disks or TDEs. It is worth noting that \cite{Liska2020} also demonstrate the development of a MAD within 1-several days, so this process is sufficiently rapid to explain the coincidence of a jet outburst with the disruption of the star.

\section{Additional Grid Details}
\label{sec:appB}

We adapt modified Kerr-Schild coordinates with the inner radius of the simulation domain inside of the BH horizon. The uniformly spaced internal coordinates $(x_1,x_2,x_3)$ are related to the Kerr-Schild spherical polar coordinates polar coordinates $(r,\vartheta,\varphi)$ by 
\begin{align}
    r &= e^{x_1},\\
    \vartheta &= \left[1 + \cot\left(\frac{H_0\pi}{2}\right) f_\vartheta(x_1)\right]\dfrac{\pi}{2},\\
    \varphi &= x_3.
\end{align}
The function in the middle expression is
\begin{multline}
    f_\vartheta(x_1)=\tan\biggl(H_0\pi\biggl[-0.5 + \left(Y_1 + \dfrac{(-Y_1 + Y_2)}{(e^{x_1}/2)^{P_0}}\right) \\  (1 - 2x_2) + x_2\biggr]\biggr).    
\end{multline}
The expression for $\vartheta$ is designed such that (i) the minimum/maximum coordinate $\vartheta$ is radially dependent, and (ii) more cells are focused towards the midplane $\vartheta=\pi/2$. We choose $H_0=0.7$ to add slightly more resolution in the midplane in order to better resolve the accretion disk. We also choose $Y_1=0.001$, $Y_2=0.02$, and $P_0=1.5$ such that $Y_2\pi<\vartheta<(1-Y_2)\pi$ near the horizon but $Y_1\pi<\vartheta<(1-Y_1)\pi$ further away. This choice ultimately increases the minimum time step and decreases the computational cost of each simulation.

\section{Initial Conditions}
\label{sec:appC}

We use the power law angular momentum disk in hydrostatic equilibrium that was presented in \citet{Kato2004} to initialize the torus in the \textsc{KORAL} code. 

For the model presented in \citet{Kato2004}, they use the pseudo-Newtonian potential described in \citet{Paczynsky1980}:
\begin{equation} \label{eq:eqA1}
  \phi = -\dfrac{GM}{(R-R_S)},
\end{equation}
where $R$ is the radius in polar coordinates, and $R_S$ is the Schwarzschild radius. A polytropic equation of state is assumed such that $p=K\rho^{1+1/n}$ and the angular momentum distribution of the disk is assumed to be a power law given by:
\begin{equation} \label{eq:eqA2}
  l(r,z) = l_0 \left(\dfrac{r}{r_0} \right)^a,
\end{equation}
where $r$ and $z$ are the cylindrical radius and height, and $l_0 = (GMr_0^3)^{1/2}/(r_0 - R_S)$. Here $r_0$ is simply a scale radius that sets the pressure and density maximum and $a$ is a constant. Under these assumptions, the condition for hydrostatic equilibrium combined with the polytropic equation of state yields a complete solution for the entire torus given the pressure ($p_0$) and density ($\rho_0$) at the characteristic radius $r_0$:
\begin{align}
  \rho &= \rho_0\left[ 1 - \dfrac{\gamma}{v_{s,0}^2}\dfrac{(\psi - \psi_0)}{n+1}\right]^{n}, \label{eq:eqA3} \\
  p &= \rho_0\dfrac{v_{s,0}^2}{\gamma}\left(\dfrac{\rho}{\rho_0}\right)^{1+1/n}, \label{eq:eqA4} 
\end{align}
where $\gamma$ is the adiabatic index (which we set to $4/3$ since the torus is radiation dominated which implies $n=3$), $v_s = \sqrt{\gamma p/\rho}$ is the sound speed of the gas, $\psi = \phi + \xi = -GM/(R-R_S) -l^2/2r^2(1-a)$ is the effective potential. Here $\xi$ is the centrifugal potential.

The Bernoulli parameter for the gas is given by the sum of the specific kinetic, potential, and internal energy. The gas is initially on a Keplerian orbit so the Bernoulli parameter is:
\begin{equation} \label{eq:eqA5}
  \rm{Be} = (1-a)\xi + \phi + \psi_{\rm{int}}, 
\end{equation}
where $\psi_{\rm{int}} = \gamma p /(\gamma - 1)\rho$ is the internal potential. The condition of hydrostatic equilibrium satisfies the equation $\nabla(\xi + \phi + \psi_{\rm{int}}) = 0$, which implies:
\begin{equation} \label{eq:eqA6}
\xi + \phi + \psi_{\rm{int}} = {\rm{constant}}.
\end{equation}
Theoretical studies of TDE disks find that the gas comes in with roughly equal angular momentum. As such, we use a constant angular momentum model in this work. This implies that we should choose $a=0$. Under this condition, the Bernoulli parameter of the disk is also constant given equations (\ref{eq:eqA5}) and (\ref{eq:eqA6}).

To initialize the disk within the \textsc{KORAL} code, we specify the characteristic radius ($r_0$), maximum density ($\rho_0$), and initial gas temperature at the density maximum ($T_0$). We set the characteristic radius $r_0=20r_g$. The initial gas density is $\rho_0=1.986\times10^{-8}\rm{g \, cm^{-3}}$ to give an initial debris cloud mass of  $\approx9.6\times10^{-3}M_\odot$ and an accretion rate roughly 10 times the Eddington mass accretion rate. The gas temperature $T_0=3.5\times10^{10}$ K is chosen such that the initial torus extends to roughly $300\,r_g$. 

To achieve a MAD accretion disk, we initialize the magnetic field as a large dipolar field by setting the vector potential scaled by the mass density 
\begin{equation}
  A_\varphi=\max\biggl( \rho r^3 \sin^3(\vartheta) \sin(r/\lambda) ,\ 0\biggr). 
\end{equation}
Here $\rho$ is specifically the gas density of the torus, which is set to $-1$ outside of the initial torus so there is no magnetic field outside of it. The wavelength $\lambda=9600\, r_g$, which is much larger than the initial torus to ensure only one sign throughout. Starting from the vector potential guarantees that the magnetic field is divergence free. We set the magnetic field strength by scaling it relative to the gas pressure with $\beta_g=10^{-3}$. This leads to the accumulation of magnetic field of only one polarization and the BH builds up a large magnetic flux quite rapidly. The peak magnetic flux in the initial torus is $\approx8\times10^{30} \rm{G \ cm^{2}}$, well above the required flux for the MAD state.

\begin{figure}
    \centering{}
	\includegraphics[width=\columnwidth]{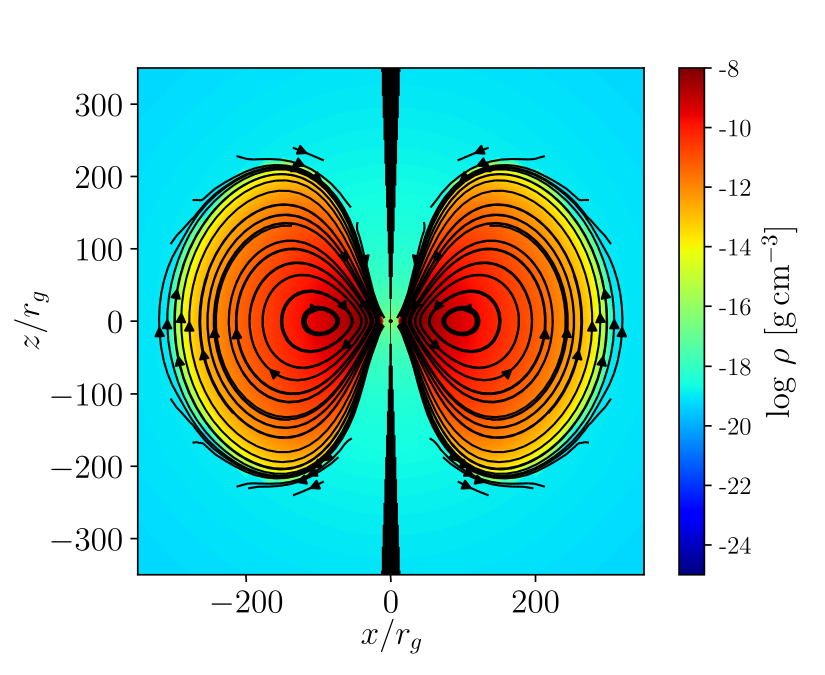}
    \caption{Here we show the gas density (colors) and magnetic field (streamlines) of the initial torus.}
    \label{fig:figA1}
\end{figure}

\begin{figure}
    \centering{}
	\includegraphics[width=\columnwidth]{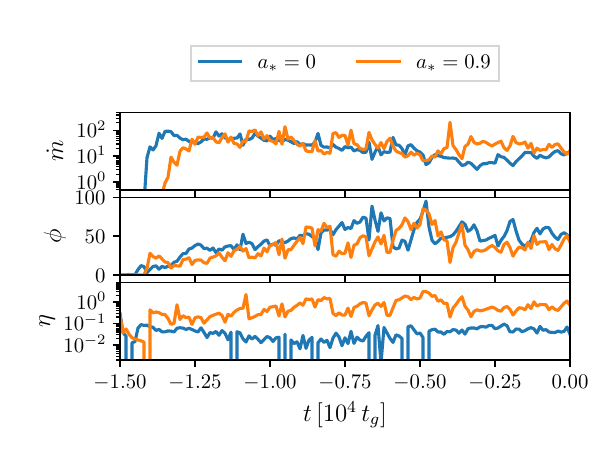}
    \caption{Here we show the initial evolution of each BH spin model prior to stream injection.}
    \label{fig:figA2}
\end{figure}

We run the initial state for a spin $a*=0$ and $a*=0.9$ for a total of $15,000t_g$. To save on resources, we run the first $10,000t_g$ in 2D $r-\vartheta$ coordinates and then regrid to the full 3D grid. When we perform the regrid, a $5\%$ perturbation is applied to the azimuthal velocity $u^\varphi$ to break the symmetry. We find that $5000t_g$ is enough for the system to become fully asymmetrical as in standard MADs. We show the gas density and field lines of the initial state in Figure \ref{fig:figA1}. We show the simulation history for both BH spins prior to injecting the TDE stream in Figure \ref{fig:figA2}.

We show the total mass injected during the stream/disk interaction phase of each simulation in Table \ref{tab:tableC1}. For the low $f_\rho$ simulations, we inject nearly as much mass as was initially in the torus. At $f_\rho>0.1$, the injected mass is quite small compared to the initial torus (10-100 times smaller).

\begin{table}
    \centering
    \begin{tabular}{ l c c c c c}
        \hline
        \hline 
        Model & $a_*$ & $\dot{M}_{\rm{inj}}$ & $t_{\rm{start}}$ & $t_{\rm{end}}$ & $M_{\rm{inj}}$ \\
                & & $(\dot{M}_{\rm{Edd}})$ & ($10^4t_g$) & ($10^4t_g$) & ($M_\odot$) \\
        \hline
        \texttt{m00f0.3b4} & 0 & 1 & 0 & 2 & $1.2\times10^{-4}$ \\
        \texttt{m00f0.003b4} & 0 & 100 & 0 & 2 & $1.2\times10^{-2}$ \\
        \texttt{m09f1b7A} & 0.9 & 1 & 0 & 2 & $4.4\times10^{-5}$ \\
        \texttt{m09f0.1b7A} & 0.9 & 10 & 0 & 3.5 & $7.8\times10^{-4}$ \\
        \texttt{m09f0.01b7} & 0.9 & 100 & 0 & 3.5 & $7.8\times10^{-3}$ \\
        \texttt{m09f1b7B} & 0.9 & 1 & 2 & 3.5 & $4.5\times10^{-3}$ \\
        \texttt{m09f0.1b7B} & 0.9 & 10 & 2 & 7 & $5.5\times10^{-4}$ \\
    \hline
    \end{tabular}
    \caption{Here we tabulate the total mass injected for each model ($M_{\rm{inj}}$) as well as relevant parameters describing the injection phase from Table \ref{tab:table1} ($\dot{M}_{\rm{inj}}$, $t_{\rm{start}}$, and $t_{\rm{end}}$).}
    \label{tab:tableC1}
\end{table}

\section{Convergence Testing}
\label{sec:appD}

\begin{figure}
    \centering{}
	\includegraphics[width=\columnwidth]{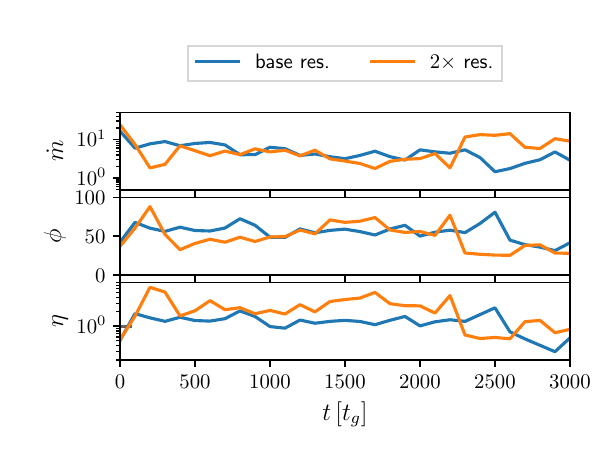}\\
	\includegraphics[width=\columnwidth]{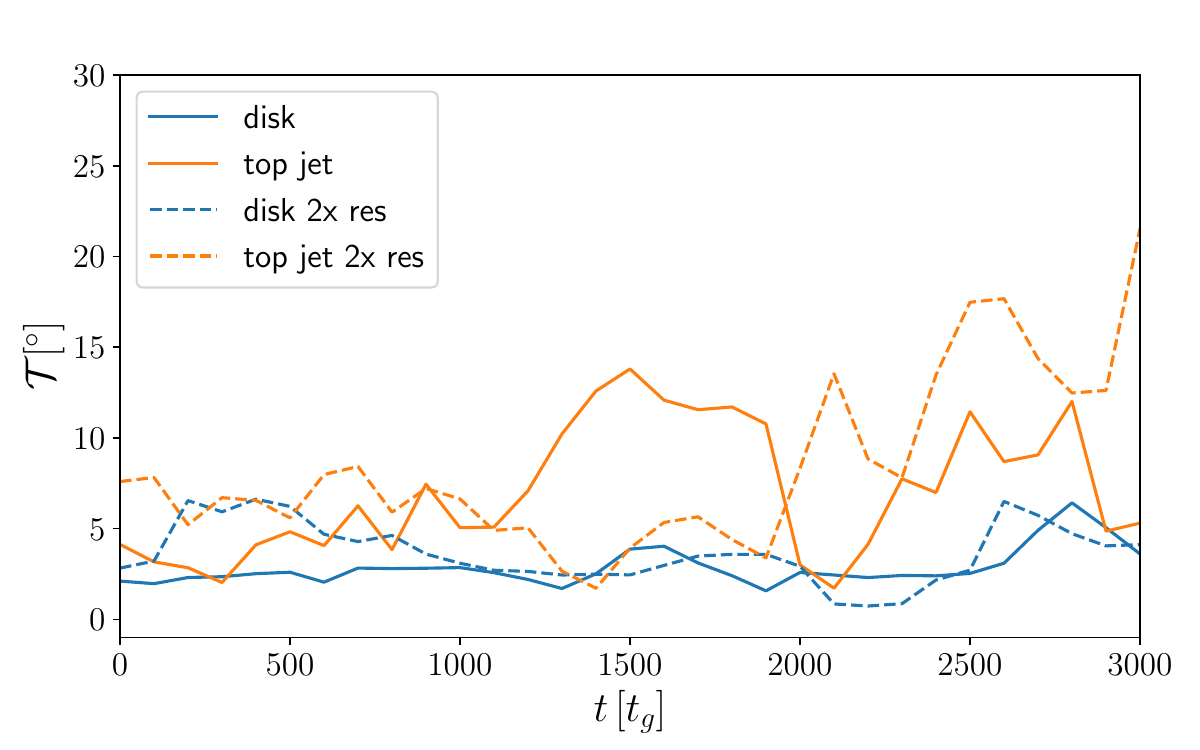}
    \caption{\textit{Top:} Here we show the initial evolution of \texttt{m09f0.01b7} at the base resolution (blue) and doubled resolution (orange) for $\dot{m}$, $\phi$, and $\eta$. \textit{Bottom:} Here we show the disk (blue) and jet (orange) tilt angles at the base resolution (solid lines) and doubled resolution (dashed lines).}
    \label{fig:figD1}
\end{figure}

We perform a simulation with the same disk and stream conditions as model \texttt{m09f0.01b7} with double the resolution in each coordinate ($N_r\times N_\vartheta\times N_\varphi=512\times288\times288$). As this simulation is extremely expensive, we only simulate the initial stream-disk interaction and self-intersection to capture the jet tilt which covers 3000 $t_g$ of data. The relevant simulation parameters for the base resolution and doubled resolution are compared in \autoref{fig:figD1}. We find excellent agreement between the two which suggests a similar degree of jet tilt is expected as we increase resolution.

A full $35,000t_g$ of convergence testing is beyond our computational capacity due to the cost of full 3D GRMHD simulations. In particular, the Courant condition near the poles severely limits the time step as more resolution is added in the azimuthal direction.

\label{lastpage}
\end{document}